\documentclass[12pt,superscriptaddress,floatfix,a4paper]{article}
\usepackage[T2A]{fontenc}
\usepackage[cp1251]{inputenc}
\usepackage[english]{babel}
\usepackage{amsmath,amssymb,amsfonts}
\usepackage{bm,latexsym,euscript,textcomp}
\usepackage{graphicx,color,graphics,caption,epsf,dcolumn}
\usepackage{natbib}
\usepackage{gensymb}
\usepackage{epstopdf}
\usepackage{booktabs}

\newcommand{\beq}{\begin{equation}}
\newcommand{\eeq}{\end{equation}}
\newcommand{\pt}{\partial}

\begin{document}

\title{\Large \bf  Spatiotemporal relationships between sea level pressure and air temperature in the tropics}

\author{A. M. Makarieva$^{1,2}$\thanks{\textit{Corresponding author.} {E-mail: ammakarieva@gmail.com}}, V. G. Gorshkov$^{1,2}$,
A.V. Nefiodov$^1$,\\ D.  Sheil$^{3,4,5,6}$, A. D. Nobre$^7$, B.-L. Li$^2$}

\date{\vspace{-5ex}}

\maketitle

\noindent
$^1$Theoretical Physics Division, Petersburg Nuclear Physics Institute, 188300 Gatchina, St. Petersburg, Russia
$^2$XIEG-UCR International Center for Arid Land Ecology, University of California, Riverside 92521-0124, USA
$^3$Norwegian University of Life Sciences, \AA s, Norway
$^4$School of Environment, Science and Engineering, Southern Cross University, PO Box 157, Lismore, NSW 2480, Australia;
$^5$Institute of Tropical Forest Conservation, Mbarara University of Science and Technology, PO Box, 44, Kabale, Uganda;
$^6$Center for International Forestry Research, PO Box 0113 BOCBD, Bogor 16000, Indonesia;
$^7$Centro de Ci\^{e}ncia do Sistema Terrestre INPE, S\~{a}o Jos\'{e} dos Campos SP 12227-010, Brazil.

\begin{abstract}
While surface temperature gradients have been highlighted as drivers
  of low-level atmospheric circulation, the underlying physical mechanisms remain unclear.  \citet{lindzen87} noted that sea level  pressure (SLP)
gradients are proportional to surface temperature gradients  if isobaric height (the height where pressure does not vary in the
  horizontal plane) is constant; their own model of low-level
  circulation assumed that isobaric height in the tropics is around 3 km. Recently \citet{bayr13} proposed a  simple model of temperature-driven
air redistribution
from  which they derived that the isobaric height in the tropics again
  varies little but occurs higher (at the height of the troposphere).
  Here investigations show that neither the empirical assumption of
  \citet{lindzen87} nor the theoretical derivations of \citet{bayr13}
  are plausible. Observations show that isobaric  height is too variable to determine a universal spatial or temporal
relationship between local values of air temperature and SLP. Since isobaric height cannot be determined from independent considerations,
the relationship between SLP and temperature is not evidence that differential heating drives  low-level circulation. An alternative
theory suggests SLP gradients are determined by the condensation of water vapor as moist air converges towards the equator. This theory
 quantifies the meridional SLP differences observed by season across the Hadley cells reasonably well. Higher temperature of surface air
where SLP is low may be determined by equatorward  transport and release of latent heat below the trade wind inversion layer.  The relationship
between atmospheric circulation and moisture dynamics merits further investigation.
\end{abstract}

\section{Introduction}

Low-level tropical winds are generally linked to convection, but the physical processes and relationships remain a matter of interest and discussion.
One question is whether the release of latent heat in the upper atmosphere generates sufficient moisture convergence in the lower atmosphere
to feed convection. The observed relationship between sea surface temperature and SLP (with warm areas having low pressure) is
regarded as evidence that low-level convergence is, rather, driven by the temperature gradients
\citep[see discussions by][]{lindzen87,neelin89,sobel06,back09,an11}.

The central concept behind surface pressure gradients driven by surface differential heating in a hydrostatic atmosphere is the existence of an
isobaric height $--$ a certain pressure level with its altitude remaining constant in either space or time despite changing surface temperature and
pressure. If such a level exists, in areas where temperature is high and air density is low there will be less air below the isobaric height
than where temperature is low and air density is high. Accordingly, surface pressure equal to the weight of the air column will be lower in
the warmer than in the colder areas. Moreover, the surface pressure and temperature gradients will be proportional to each other, with the
proportionality coefficient set by the isobaric height.

The main problem is to find the isobaric height. While for a liquid a natural candidate is the height of the upper surface, our gaseous atmosphere
lacks a sharp upper boundary. \citet{lindzen87} suggested that isobaric height in the tropics approximately coincides with the height of the trade
wind inversion ($\approx 3$~km). However, no theoretical justifications were offered. More recently \citet{bayr13} suggested the height of the troposphere
could be isobaric. Specifically, \citet{bayr13} proposed a simple physical model where air is re-distributed between air columns subjected to differential heating so
as the height of a certain pressure level remains the same. From this model \citet{bayr13} concluded that this constant height is the height of the
troposphere at $H \approx 16$~km and that it determines the proportionality between changes of the mean tropospheric temperature and SLP.

So is the tropical isobaric height $3$~km following \citet{lindzen87} or 16~km following \citet{bayr13}?
If we knew that a constant isobaric height exists in the tropics and could link its magnitude to known atmospheric parameters,
we could use surface temperatures to predict surface pressures and, resulting from air circulation, moisture convergence and precipitation.
Our incomplete understanding of the physical principles governing low-level circulation is manifested by the inability of atmospheric models to
replicate the terrestrial water cycle \citep{ma06,hagemann11} as well as by the challenge of confidently predicting precipitation and air
circulation under a changing climate \citep[e.g.,][]{an11,huang13}.

To explore the relationship between surface pressure and temperature we start by re-examining the derivation of \citet{bayr13}.
We identify and resolve several inconsistencies in their model (Section~\ref{bayr}). We demonstrate that the height of the tropical troposphere
is not an isobaric height and that it does not determine the ratio between the mean tropospheric temperature and SLP changes. We then derive a general relationship linking the ratio of gradients
(as well as of temporal changes) of surface pressure and temperature to an isobaric height. We show that this ratio is a function not of one but of two heights, isobaric and isothermal (Section~\ref{ih}).

Using data provided by the National Centers for Environmental Prediction/National Center for Atmospheric Research (NCEP/NCAR) Reanalysis \citep{kalnay96} and the Remote Sensing Systems \citep{mears09} we investigate the spatial and temporal relationships between tropical
SLP and air temperature on a seasonal timescale (Section~\ref{data}).
These data show no constant isobaric height in the tropical atmosphere. The isobaric height (defined as the height where the meridional pressure gradient is zero) in fact grows sharply from zero around the 30th latitudes to above the tropospheric height at the equator.
We show that the isobaric height varies with location and season.

We then demonstrate that in theory a constant isobaric height dictates a proportionality between pressure gradients at the surface and in the upper atmosphere irrespective
of whether the latter are driven by differential heating at the surface or in the upper atmosphere. We demonstrate that in the real atmosphere such a proportionality does not exist
(Section~\ref{sdiff}). We argue that the mere existence of a relationship between surface gradients of pressure and temperature
does not by itself imply causality and is thus insufficient to conclude that surface pressure gradients are driven by differential heating.

We here propose an alternative concept to understand and quantify the observed surface pressure variation in the tropics.
Horizontal transport of moisture with its subsequent condensation and precipitation away from the point where it evaporated produces pressure
gradients due to the changing air concentration as it moves from the evaporation to condensation area \citep{m13,mgn14}. Pressure is greater
where water vapor is added and lower where it is removed from the air column.
Horizontal moisture transport thus appears to be the direct cause of the surface pressure gradients that, in their turn, maintain this transport
and the associated convection. This theoretical approach effectively describes the seasonal dynamics of surface pressure differences across the Hadley
cells (Section~\ref{vapor}).
Finally, we show that the horizontal transport of moisture can also account for the association between the surface pressure and temperature gradients (Section~\ref{temp}).

\section{The physical model of \citet{bayr13}}
\label{bayr}

\citet{bayr13}  begin their derivation with an equation they refer to as "the hydrostatic equation"
\beq\label{b1}
dp = -\rho g d\eta
\eeq
with pressure $p$, density $\rho$, gravity constant $g$, and $\eta$
described as "air column height"\footnote{In the derivation of \citet{bayr13} $\eta$ in (\ref{b1}) is denoted as $h$.}.
According to \citet{bayr13}, for an "isobaric thermal expansion of the air column"
it follows from the ideal gas law that
\beq\label{b2}
d\eta = \frac{\eta}{T}dT,
\eeq
where $T$ is temperature.
\citet{bayr13} propose that "to balance the heights of the two columns at the end,
half of the height difference is moved from the warmer to the colder air volume". They conclude that using
Eqs.~(\ref{b1}) and (\ref{b2}) one obtains how SLP depends on temperature
\beq\label{b3}
\frac{dp}{dT} = \frac{1}{2}\rho g \frac{\eta}{T}.
\eeq
We first note
that the resulting equation (\ref{b3}), which forms the basis for all analyses presented by \citet{bayr13},
mathematically contradicts the preceding
equations (\ref{b1}) and (\ref{b2}) from which it presumably derives.
Indeed, combining (\ref{b1}) and (\ref{b2}) we find that the minus sign in (\ref{b3}) has been lost, while the
$1/2$ multiplier has been added (cf. Eq.~(\ref{ze}) below).

The physical validity of the derivation (\ref{b1})-(\ref{b3})
is further undermined by the use of the incorrect hydrostatic equation (\ref{b1})
and by the lack of an explicit definition of the key variables.
For atmospheric air conforming to the ideal gas law
\beq\label{ig}
p = N R T,\,\,\,R = 8.3~{\rm J~mol^{-1}~K}^{-1},
\eeq
where $N$ is molar density, the correct hydrostatic equilibrium equation is
\beq\label{he}
dp(z) = -\rho(z) g dz,\,\,\,\frac{\pt p}{\pt z} = -\frac{p}{h},\,\,\,h\equiv \frac{RT}{Mg},
\eeq
where $M$ is molar mass. Here $z$ is not the "air column height", but an arbitrary height in the atmosphere, $dp(z)$ is not the change
of SLP with time, but the spatial change of air pressure over a small vertical distance $dz$ at height $z$.
Note that the exponential scale height $h$ of pressure conforms to (\ref{b2}) but not to (\ref{b1}).

The "air column height" $\eta$ is never formally defined
by \citet{bayr13} despite balancing this particular height is key to their model.
To estimate $dp/dT$ from (\ref{b3}) \citet{bayr13} set $\rho$ in (\ref{b3})
equal to the mean air density in the troposphere $\rho = \rho_a = 0.562$~kg~m$^{-3}$.
They take $\eta$ equal to the height $H$ of the tropical troposphere $\eta = H = 16.5$~km and temperature $T$
equal to the mean tropospheric temperature $T_a$ defined as the mean air temperature below 100 hPa,
$T = T_a = 263.6$~K. Then (\ref{b3}) gives $dp/dT = 1.7$~hPa~K$^{-1}$.

Density $\rho$ in (\ref{b3}) and (\ref{b1}) is not well defined either as it is not specified to
which part of the atmospheric column it pertains.
While air temperature $T(z)$ varies by about 30\% at most from its surface value to the top of
the troposphere, tropospheric air density $\rho(z)$ as well as air pressure $p(z)$ vary by an order of magnitude.
In their quantitative estimate \citet{bayr13} interpreted $dp$ as describing {\it sea level} pressure change
but for density $\rho$ they took the {\it mean tropospheric} instead of surface value.
This choice was incorrect as we demonstrate below (see Eq.~(\ref{ze}) in the next section).

\section{Isobaric height}
\label{ih}

The model of \citet{bayr13} did not consider how temperature might vary with height.
We will here derive a general relationship linking surface pressure and
temperature to the vertical structure of the atmosphere. We will allow air temperature
to vary with height with a lapse rate $\Gamma \equiv -\pt T/\pt z$, which is
independent of height but can vary in the horizontal direction.

We introduce the following dimensionless variables to replace height $z$ and lapse rate $\Gamma$:
\beq\label{not}
Z \equiv \frac{z}{h_s},\,\,\,c \equiv \frac{\Gamma}{\Gamma_g},\,\,\,h_s \equiv \frac{RT_s}{Mg} \equiv \frac{T_s}{\Gamma_g},\,\,\,
\Gamma_g \equiv \frac{Mg}{R}=34\,{\rm K\,km}^{-1},\,\,\,M = 29~{\rm g~mol}^{-1}.
\eeq
For air temperature we have
\beq\label{T}
T(Z) =T_s(1 - c Z),\,\,\,Z < c^{-1},\,\,\,T_s \equiv T(0).
\eeq
The hydrostatic equilibrium equation (\ref{he}) assumes the form
\beq\label{heG}
-\frac{\pt p}{\pt Z} = \rho g h_s = \frac{p}{1-c Z}.
\eeq
Solving (\ref{heG}) for $p \ge 0$ we have
\beq\label{heG1}
\ln\frac{p}{p_s} = -\int_0^Z \frac{dZ'}{1-c Z'}= \frac{1}{c}\ln(1-c Z) \approx
- Z - \frac{1}{2}c Z^2.
\eeq
The approximate equality in (\ref{heG1}) holds for $c Z \ll 1$, which corresponds to $z \ll h_s (\Gamma_g/\Gamma) = 45$~km,
which is always the case in the troposphere.

Pressure $p(z)$ and temperature $T(z)$ at a given height $z$ are functions of $p_s$, $T_s$ and $\Gamma$. Taking the total differential
of the approximate relationship for $p$ (\ref{heG1}) over these three variables we obtain:
\beq\label{dp}
dp = p_s (da + Zdb - \frac{1}{2}Z^2dc) e^{-Z},
\eeq
where $da$, $db$ and $dc$ stand for the dimensionless differentials of $p_s$, $T_s$ and $\Gamma$:
\beq\label{diff}
da \equiv \frac{dp_s}{p_s} \approx \frac{dp_s}{\overline{p_s}},\,\,\,
db \equiv \frac{dT_s}{T_s}\approx \frac{dT_s}{\overline{T_s}},\,\,\,
dc \equiv \frac{d\Gamma}{\Gamma_g},
\eeq
where $\overline{p_s} = 1013$~hPa is the annual mean SLP, $\overline{T_s} = 298$~K
is the annual mean surface air temperature in the tropics.
With height $z$ fixed, these differentials describe the change of respective variables in time and/or horizontal dimension.
The inaccuracy of the approximate relationships in (\ref{diff}) is determined by the
relative changes of SLP and surface temperature across the tropics. For the zonally averaged $p_s$
and $T_s$ this inaccuracy does not exceed $4\%$.

Isobaric height $z_e \equiv Z_e h_s$ is defined from (\ref{dp}) as the height where $dp = 0$.
It is determined from the following quadratic equation:
\beq\label{Ze}
da + Z_e db - \frac{1}{2} Z_e^2 dc = 0, \,\,\,Z_e = \frac{db}{dc}\left(1 \pm \sqrt{1+2 \frac{da}{db}\frac{dc}{db}}\right).
\eeq
There are at maximum two isobaric heights. Note that the isobaric height $Z_e$ (\ref{Ze}) does not depend on lapse
rate $c$ but only on its differential $dc$. This is a consequence
of the smallness of $c Z \ll 1$ in the troposphere.

When $db = 0$, i.e., when the surface temperature does not vary, but only lapse rate does,
we have from (\ref{Ze})
\beq\label{a/c}
\frac{da}{dc} = \frac{Z_e^2}{2}.
\eeq
The surface pressure change is proportional to the change in lapse rate, i.e. the pressure is lower
where the lapse rate is smaller, with the proportionality coefficient equal to half the squared isobaric height.

By analogy with the isobaric height, isothermal height $z_i \equiv Z_i h_s$ is found by taking
total differential of $T$ (\ref{T}) over $T_s$ and $\Gamma$ and putting $dT = dT_s - T_s Z_i dc = 0$. This gives
\beq\label{Zi}
Z_i = \frac{db}{dc} = \frac{1}{h_s}\frac{dT_s}{d\Gamma},\,\,\,z_i \equiv Z_i h_s = \frac{dT_s}{d\Gamma}.
\eeq

From (\ref{Ze}) and (\ref{Zi}) we obtain the following relationship for the
ratio of the differentials of surface pressure and temperature (\ref{diff}):
\beq\label{a/b}
\frac{da}{db} = -Z_e \left(1-\frac{1}{2} \frac{Z_e}{Z_i}\right).
\eeq

When, as in the model of \citet{bayr13}, lapse rate is assumed to be constant with $dc = 0$, we have $Z_i = \infty$
and (\ref{a/b}) becomes (cf. \ref{a/c})
\beq\label{Ze1}
\frac{da}{db} = -Z_e.
\eeq
Expressing this result using notations (\ref{diff}) we find
\beq\label{ze}
\frac{dp_s}{dT_s}=-\frac{z_e}{h_s}\frac{p_s}{T_s} = -\rho_s g\frac{z_e}{T_s}.
\eeq
Comparing (\ref{ze}) to (\ref{b3}) of \citet{bayr13} we notice the absence of coefficient $1/2$ in (\ref{ze})
and the presence of surface air density $\rho_s$ in (\ref{ze}) instead of an undefined air density $\rho$ in (\ref{b3}).
If, as did \citet{bayr13}, one assumes $z_e$ to be equal to the tropospheric height
$H = 16.5$~km, then Eq.~(\ref{ze}) with $\rho_s = 1.2$~kg~m$^{-3}$ and $T_s = 300$~K yields
$dp_s/dT_s = -6.5$~hPa~K$^{-1}$. This estimate is 2.7 times greater by absolute magnitude than the ratio $dp_s/dT_a =-2.4$~hPa~K$^{-1}$
obtained by \citet[][their Fig.~2]{bayr13} from observations
(note that in the case considered in the model of \citet{bayr13} with $dc = 0$ we have $dT_s/dT_a = T_s/T_a \approx 1$
and $dp_s/dT_a \approx dp_s/dT_s$, see Appendix~\ref{app}).
This discrepancy shows that neither the tropospheric height nor the mean tropospheric density determine the ratio of pressure and
temperature changes in the tropical atmosphere.

In the general case the ratio $da/db = (dp_s/dT_s)(T_s/p_s)$ (\ref{a/b}) is controlled not only by the isobaric height
$Z_e$ but also by the isothermal height $Z_i$.  Ratios $da/db$ and $db/dc$ in (\ref{Ze}) and (\ref{Zi})
can be understood as the ratios of the gradients of the corresponding variables, e.g.
$da/db = (\pt p_s/\pt y)/(\pt T_s/\pt y)(T_s/p_s)$, where $(\pt p_s/\pt y)/(\pt T_s/\pt y)$
is the ratio of pressure and temperature gradients in a given $y$ direction (e.g. along the meridian). In this case for any $y$ the
value of $z_e$ (or $z_i$) has the meaning of a height where  $\pt p/\pt y = 0$ (or $\pt T/\pt y = 0$), i.e. where pressure
(or air temperature) does not vary over $y$.
Second, for a particular grid point these ratios can be understood as the ratio of temporal derivatives:
$dp_s/dT_s = (\pt p_s/\pt t)/(\pt T_s/\pt t)$. In this case $z_e$ and $z_i$ represent the height of a pressure level
and a temperature level that do not change with time.
Finally, these ratios can be understood as the ratios of small finite differences between pressure
or temperature in a given grid point and a certain reference value of pressure or temperature,
$dp_s/dT_s = \Delta p_s/\Delta T_s$.

In all these cases the proportionality between pressure and temperature variations, either temporal or spatial, will result
if $z_e$ and $z_i$ are constant, see (\ref{a/b}). We can estimate all parameters in (\ref{a/b}) from empirical data to see
if such a relationship holds across the tropics.

\section{Data analysis}
\label{data}

We used NCAR-NCEP reanalysis data on SLP and surface air temperature, as well as on geopotential height and air temperature at
13 pressure levels provided by the NOAA/OAR/ESRL PSD, Boulder, Colorado, USA, from their Web site at {\it http://www.esrl.noaa.gov/psd/} \citep{kalnay96}.
As an estimate of the mean tropospheric temperature we took TTT (Temperature Total Troposphere) MSU/AMSU satellite data
provided by the Remote Sensing Systems from their Web site at {\it http://www.remss.com/measurements/upper-air-temperature} \citep{mears09}.
Monthly values of all variables were averaged over the time period from 1978 (the starting year for the TTT data) to 2013
to obtain 12 mean monthly values and one annual mean for each variable for each grid point on a regular $2.5\degree \times 2.5\degree$
global grid.\footnote{TTT data array contains 144 (360/2.5) longitude and 72 (180/2.5) latitude values each pertaining to
the center of the corresponding grid point. NCAR-NCEP data arrays contain 144 longitude and 73 latitude values each pertaining to the border
of the corresponding grid point. E.g., the northernmost latitude in the NCAR-NCEP data is 90$\degree$N, while for the TTT data it is
$90 - 2.5/2 = 88.75\degree$N. This discrepancy was formally resolved by adding an empty line to the end of the TTT data
such that the number of lines match and matching $i, j$ grid points in the two arrays. In the result, every TTT value
refers to a point in space that is 1.25 degree to the South and to the East from the coordinate of the corresponding
NCAR-NCEP value. This relatively small discrepancy did not appear to have any impact on any of the resulting quantitative conclusions (i.e. if instead
one moves TTT points to the North, the results are unchanged).}

All variables were zonally averaged. Meridional gradients $\pt X/\pt y$ of variable $X$ ($X = p_s,\, T_s$) at latitude $y$ were determined as
the difference in $X$ values at two neighboring latitudes and dividing by 2.5$\degree$:
$\pt X(y)/ \pt y \equiv [X(y+1.25\degree)-X(y-1.25\degree)]/2.5\degree$.
Meridional pressure gradients corresponding to pressure level $p_j$ were calculated from the geopotential height gradient
$\pt p_j/\pt y = (\pt z_j /\pt y) p_j/h_j$, where $z_j$ is the geopotential height of pressure level $p_j$, $h_j = RT_j/(Mg)$ is the
exponential pressure scale height (\ref{he}) and $T_j$ is air temperature at this level.
The following pressure levels covering the tropical troposphere were considered:
1000, 925, 850, 700, 600, 500, 400, 300, 250, 200, 150, 100 and 70 hPa.
Isobaric height $z_e$ at each latitude was determined as the minimal height where the meridional pressure gradient
changes its sign.

\begin{figure*}[h]
\centerline{
\includegraphics[width=0.95\textwidth,angle=0,clip]{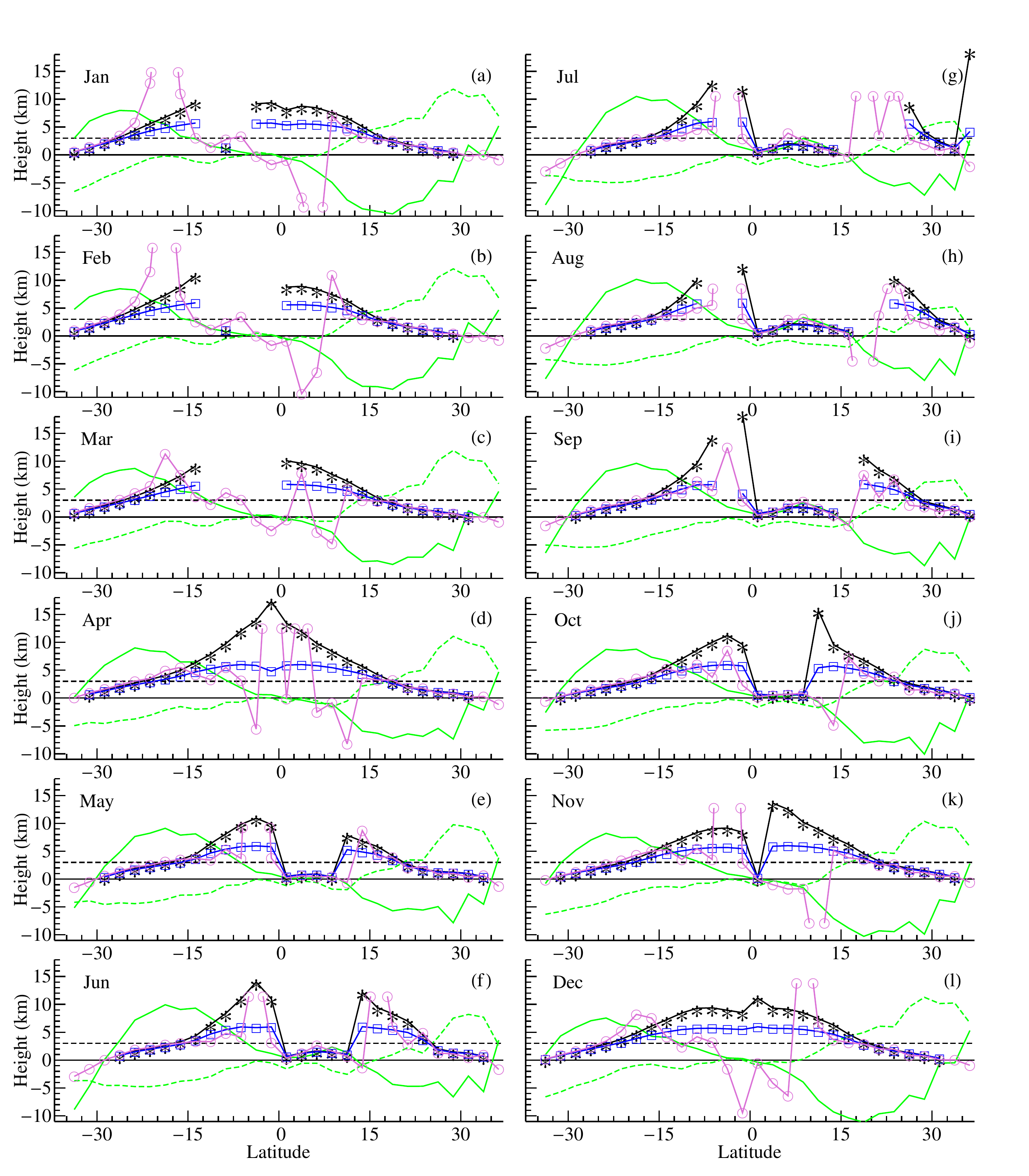}
}
\caption{\label{figze}
%\large
The observed isobaric height $z_e \equiv Z_e h_s$ (black curve with asterisks) (the height where the meridional
pressure gradient $\pt p(z_e)/\pt y = 0$),
the observed ratio of the meridional gradients of SLP and surface air temperature $-(da/db)h_s
= -(\pt p_s/\pt y)/(\pt T_s/\pt y)(T_s/p_s)h_s$
(purple curve with open circles) and its
theoretical estimate (\ref{Ze}) with isothermal height $z_i \equiv Z_i h_s = 12$~km (blue curve with open squares).
Missing points indicate latitudes where the meridional pressure gradient does not change its sign anywhere
between the 1000 hPa and 70 hPa pressure levels.
Green curves show (minus one times) the meridional gradients of SLP (solid, in $0.05$~hPa~$(\degree\rm lat)^{-1}$) and surface air
temperature (dashed, in $0.1$~K~$(\degree\rm lat)^{-1}$). The negative values are used to ease readability.
Note that the sharp fluctuations in the purple curve correspond to
latitudes where the surface temperature gradient is near zero ($db = 0$).
}
\end{figure*}

In Fig.~\ref{figze} we plotted the observed isobaric height $z_e$ and
compared it with the observed ratio of the meridional gradients of SLP
and surface air temperature $-h_s (da/db) \equiv -h_s (dp_s/dT_s)(T_s/p_s) = -h_s (\pt p_s/\pt y)/(\pt T_s/\pt y)(T_s/p_s)$.
There are two take-away messages from Fig.~\ref{figze}. First, the isobaric height of the tropical
atmosphere is not constant:
it rises steeply from zero at the outer borders
of Hadley cells to above the top of the troposphere near the equator. During some months (e.g., June, July, August) it also has a trough
at the equator. Second, the isobaric height does not universally determine the local ratio between surface gradients of pressure and temperature
as illustrated by the discrepancy between the purple and black curves.
The two curves have a tendency to match at low and depart from one another at high values of empirical $z_e$.
The observed meridional variation of $z_e$ is associated with the variation in the direction of geostrophic zonal winds.
Since the velocity of these winds is proportional to the meridional pressure gradient,
at $z = z_e$ they have zero velocity. The surface $z = z_e$ is the surface where zonal winds change their direction
\citep[cf. Fig.~1a of][]{sc06}. Beneath this roof-like surface (with slopes in the two hemispheres) the
pressure falls towards the equator and the zonal winds blow from East to West.

\begin{figure*}[h]
\centerline{
\includegraphics[width=0.90\textwidth,angle=0,clip]{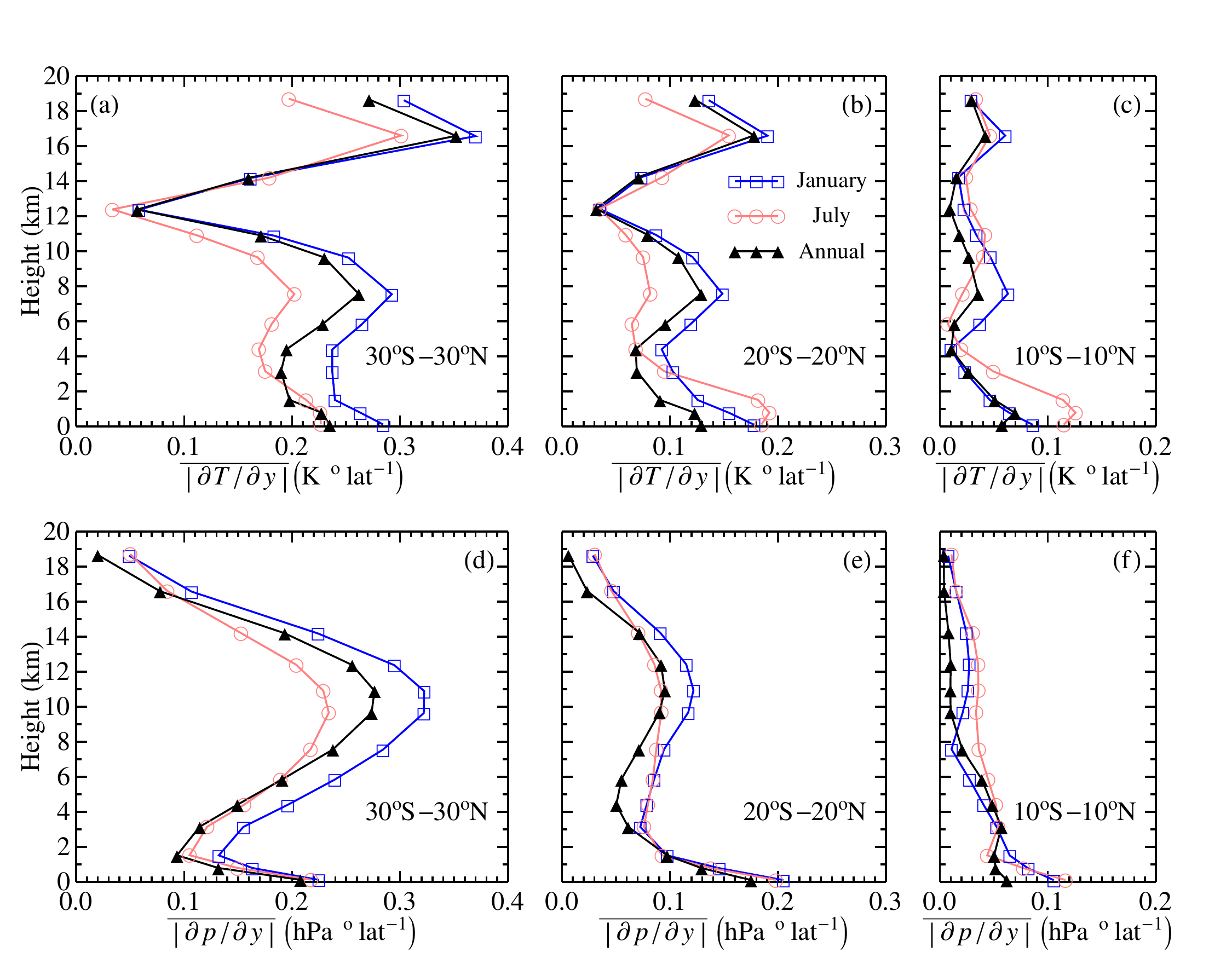}
}
\caption{\label{figtemp}
%\large
Vertical profiles of the meridional temperature (a-c) and pressure (d-f) gradients
taken by absolute magnitude and averaged from $30\degree$S to $30$$\degree$N (a,d),
$20\degree$S to $20\degree$N (b,e) and $10\degree$S to $10\degree$N (c,f) in January (blue squares), July (pink circles)
and annually (black triangles). The pantropical constant isothermal height $z_i \approx 12$~km corresponds to a minimum of
$\overline{|\pt T/\pt y|}$ that is practically independent of the averaging area. In contrast, height of the minimum
pressure gradient in the lower atmosphere moves upwards from less than 2~km (d) to 8-10~km (f) as the averaging
area decreases. This reflects the growth of isobaric height towards the equator (cf. Fig.~\ref{figze}a,g).
}
\end{figure*}

To estimate the observed $da/db$ ratio from (\ref{a/b}) we need to know the isothermal height $Z_i$.
In their model \citet{lindzen87}
adopted a constant isothermal height equal to 10~km. They observed that the horizontal temperature
differences at the level of $z_{LN} = 3$~km are 30\% smaller than the corresponding differences
at the sea level: $\Delta T(z_{LN}) = 0.7 \Delta T_s$. From $T(z_{LN}) = T_s - \Gamma z_{LN}$ and (\ref{Zi}) we obtain
$z_i \equiv \Delta T_s/\Delta \Gamma = z_{LN}/0.3 = 10$~km.
This estimate is in approximate agreement with observations of the zonally averaged temperature gradient:
the isothermal height in the tropical atmosphere corresponds to the pressure level of 200 hPa or about 12~km (Fig.~\ref{figtemp}).
The blue curve in Fig.~\ref{figze} shows that Eq.~(\ref{a/b}) describes the observed $da/db$ ratio better than Eq.~(\ref{Ze1}) (black curve)
(note that in the regions where $db \approx 0$ and $da/db$ apparently cannot be estimated from the data with sufficient accuracy).

Taking a derivative of $-da/db$ (\ref{a/b}) over latitude $y$  at constant $Z_i$
\beq\label{abd}
\frac{\pt}{\pt y}\left(-\frac{da}{db}\right) = \frac{\pt Z_e}{\pt y}\left(1 - \frac{Z_e}{Z_i}\right)
\eeq
reveals that the $-da/db$ ratio has an extremum (maximum) for $Z_e =Z_i$, i.e.
where the isobaric height $z_e$ approaches 12 km. With $Z_e$ growing beyond $Z_i$ ($\pt Z_e/\pt y > 0,\,Z_e > Z_i$), the derivative
changes its sign and $-da/db$ starts to decline.  As the term in brackets in (\ref{abd}) is less than unity, the meridional variation of $-da/db$ ratio is always less
than that of the isobaric height $Z_e$. When $Z_e$ reaches twice the isothermal height, $Z_e \to 2Z_i$,
from (\ref{a/b}) we have $-da/db \to 0$. This is a point of singularity, with $da = db = dc = 0$ and pressure
coinciding between the considered equatorial columns at all heights, including $z=0$ and $z=z_i$ (see the green line
in Fig.~\ref{figdiff}e,h below).

This equatorial minimum of $-da/db$ is relevant to the problem of "back pressure" in the model of \citet{lindzen87}.
\citet{lindzen87} proposed that pressure differences are negligible along height $z_{LN} = 3$~km which corresponds
to pressure level of 700 hPa. The $-da/db$ ratio corresponds to this height
around the 20th latitudes where the absolute magnitude of the pressure gradient is the largest (Fig.~\ref{figze}). At lower
latitudes it grows to about five kilometers to decline to near zero in the immediate vicinity of the equator in some months.
If $-da/db$ ratio is assumed to be a constant corresponding to $z_e = 3$~km, this leads to an overestimate
of the pressure gradient near the equator and an overestimate of the equatorial moisture convergence.
To cope with this problem \citet{lindzen87}
introduced a "back pressure" correction to their model which adjusted the near-equatorial pressure field to fit the observations.
However, we can see from Fig.~\ref{figze} that the concept of a constant isobaric height linking surface
pressure and temperature does not hold at large in the tropics.
In particular, the assumption of \citet[][their Eq.~9a]{lindzen87} that the latitudinal variation in $z_e$ (or $-da/db$) is small
 apparently does not
hold\footnote{We make a brief comment on an atmosphere where as in the model of
\citet{lindzen87} the isobaric height would be constant. How would winds depend on $z_e$ in such an atmosphere?
A small isobaric height at fixed surface temperature gradients means that the surface pressure gradients are small.
In the limit $z_e \to 0$ the surface pressure gradients disappears and the low-level winds should vanish.
Contrary to this expectation \citet{lindzen87} found little dependence of meridional winds on $z_e$ in their model.
A smaller $z_e$ expectedly produced weaker surface pressure gradients, but
it also produced a proportionally larger damping coefficient $\epsilon \equiv C_D |V_c|/z_e$, where $C_D$ is a constant
and $V_c$ is a typical wind speed at $z_e$ taken by \citet{lindzen87} to be equal to 8~m~s$^{-1}$.
As a result of a weaker meridional pressure gradient, zonal wind did decrease proportionally to the surface pressure gradient. However, the meridional
wind proportional to the product of zonal wind and the damping coefficient $\epsilon$ \citep[][see their Eq.~12a]{lindzen87}, did not change much.
The decrease in pressure gradient was offset by an increase in the damping coefficient $\epsilon$,
such that the low-level air convergence remained approximately independent of $z_e$.
However, this conclusion critically derives from the assumed constancy of $V_c$ $--$ characteristic wind speed at the top of the boundary layer.
In reality $V_c$ is not independent of $z_e$. In the model of \citet{lindzen87} the boundary layer height was assumed to be equal to $z_e \sim 3$~km,
which is unrealistic. In the real atmosphere the height of boundary layer $h_b$ is much smaller, $h_b \sim 1\,{\rm km} \ll z_e$.
Because of this, pressure gradients at the top of the boundary layer are determined by the surface pressure gradients and close to them.
Since at the top of the boundary layer winds are approximately geostrophic \citep{back09},
this means that the geostrophic wind speed $V_c$ at the top of the boundary layer (which is used in the determination of the damping coefficient)
is approximately proportional to the surface pressure gradient. Consequently, it decreases with decreasing $z_e$. In the result, with decreasing $z_e$
(decreasing surface pressure gradient), surface winds will decline as well proportionally to the declining $V_c$.}.

Another illustration to the same problem is provided by the results of \citet{bayr13}.
\citet[][their Fig.~2]{bayr13} made a regression of spatial SLP differences
$\Delta p_s \equiv p_s - \overline{p_s}$ versus mean tropospheric temperature differences
$\Delta T_a \equiv T_a - \overline{T_a}$, where $p_s$ and $T_a$ are values in a given gridpoint and $\overline{p_s}$ and
$\overline{T_a}$ are the mean tropical values\footnote{Note that Fig.~2 of \citet{bayr13} describes the
relationship between spatial differences of pressure and temperature rather than between their temporal changes.
In that figure $\Delta p_s$ and $\Delta T_a$ values from the four seasons are plotted together. It is clear that if there were no seasonal change
of $\Delta p_s/\Delta T_a$ whatsoever, such a regression would nevertheless produce a non-zero slope reflecting the time-invariable spatial association between
higher temperature and lower pressure.}. By construction, this regression
line goes through the axis origin ($\Delta p_s =0,\,\Delta T_a =0$).
The regression slope of $\Delta p_s/\Delta T_a = -2.4$~hPa~K$^{-1}$ obtained
by \citet{bayr13} corresponds to $\Delta p_s/\Delta T_s \approx -1.3$~hPa~K$^{-1}$ (see Appendix~\ref{app} on
the relationship between the mean tropospheric temperature $T_a$ and surface temperature $T_s$).
From (\ref{ze}) for $T_s = 298$~K,
$p_s = 1013$~hPa and $h_s = 8.7$~km (\ref{not}) and $\Delta p_s/\Delta T_s = -1.3$~hPa~K$^{-1}$ we obtain
an average $z_e = 3$~km in agreement with the assumption of \citet{lindzen87}.
However, as the linear
regression minimizes the departure
of the empirical points from the theoretical curve,
the slope of a regression line
that goes through the axes origin is set by the values that depart most
from the zero point.
The smaller $\Delta p_s$ and $\Delta T_s$ values
make the least contribution to the determination of the regression slope.
Therefore, the regression made by \citet{bayr13} does not actually estimate the pantropical mean value of the ratio between pressure and temperature
variations. Rather, the regression slope characterizes the value of this
ratio where $\Delta p_s$ and $\Delta T_s$ are the largest.

Our own analysis of the seasonal dynamics of the relationship between pressure and temperature
confirms the absence of a universal ratio between pressure and temperature changes.
For each grid point, we made a reduced major axis regression of the monthly changes of pressure $\widetilde{\Delta} p_s$ on
the monthly changes of temperature $\widetilde{\Delta} T_s$. Here
 $\widetilde{\Delta} p_s\equiv p_s(m_2) - p_s(m_1)$ and $\widetilde{\Delta} T_s\equiv T_s(m_2) - T_s(m_1)$, where $m_1$
and $m_2$ are two consecutive months (e.g., December and January).
A similar analysis was performed for $p_s$ and $T_a$.

\begin{figure*}[h]
\centerline{
\includegraphics[width=0.95\textwidth,angle=0,clip]{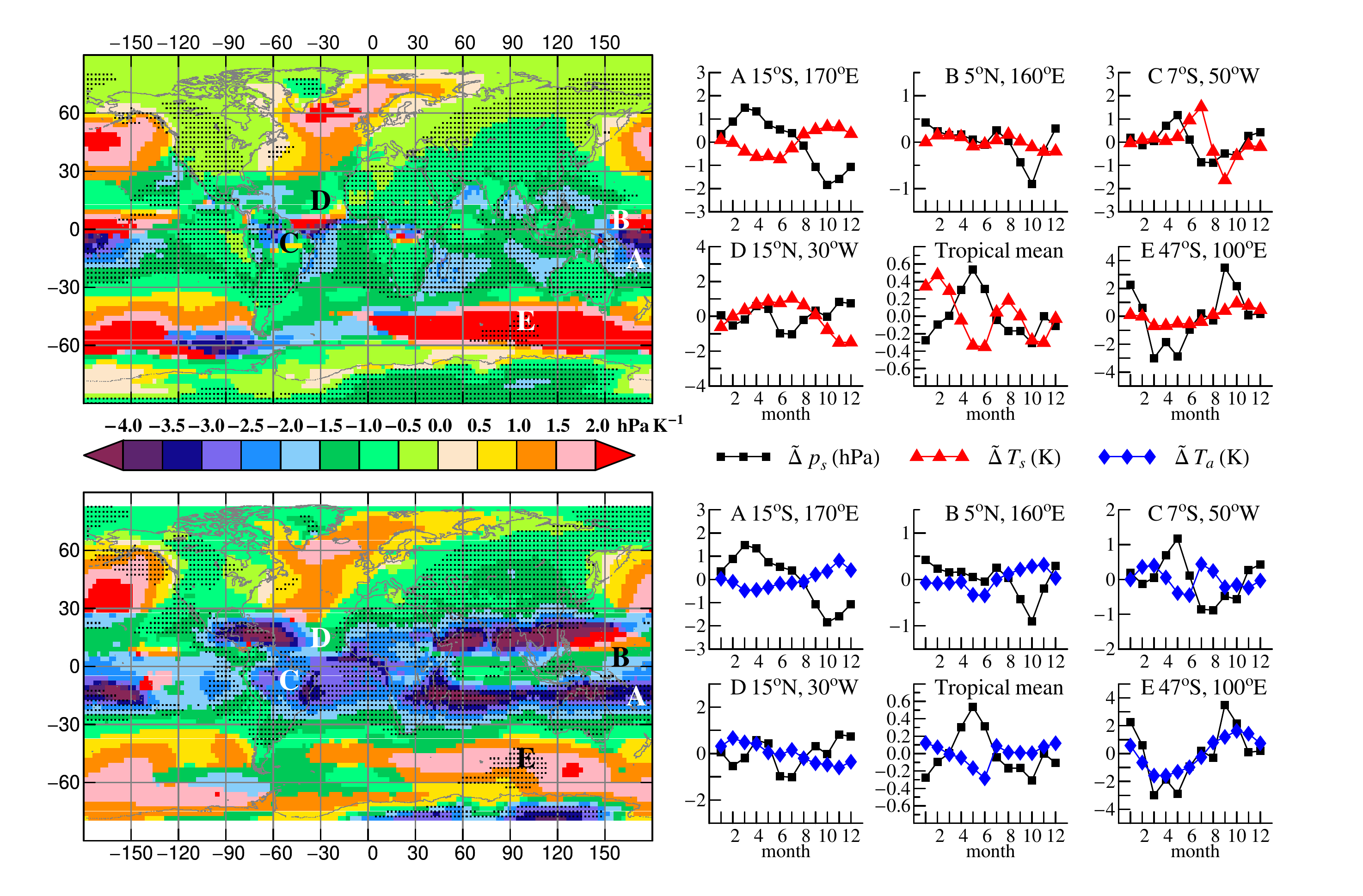}
}
\caption{\label{figmap}
%\large
Mean ratio between local monthly changes of SLP $p_s$ and surface temperature $T_s$ (larger top left panel) and
SLP and tropospheric temperature $T_a$ (larger lower left panel). The ratio is estimated as the slope coefficient of a Reduced Major
Axis regression of $\widetilde{\Delta} p_s \equiv p_s(m_2) - p_s(m_1)$ on, respectively,
$\widetilde{\Delta}T_s \equiv T_s(m_2) - T_s(m_1)$ and $\widetilde{\Delta} T_a \equiv T_a(m_2) - T_a(m_1)$,
where $\widetilde{\Delta}p_s$, $\widetilde{\Delta}T_s$ and $\widetilde{\Delta}T_a$ are the monthly changes of the respective variables
between two consecutive months $m_1$ and $m_2$.
Black dots indicate where the probability level of the regression is less than $0.01$.
The small panels exemplify seasonal changes of $\widetilde{\Delta}p_s$, $\widetilde{\Delta}T_s$ and $\widetilde{\Delta}T_a$
in individual grid points (A, B, C, D and E) shown in the big panels, as well as the tropical mean (the area between 27.5$\degree$S
and 27.5$\degree$N). Note the different vertical scales in the small panels. Tropical mean ($\pm$ standard deviation) of the obtained local slope
coefficients  are
$-1.1\pm 1.0$~hPa~K$^{-1}$ (land $-0.98\pm 0.62$~hPa~K$^{-1}$, ocean $-1.09\pm 1.11$~hPa~K$^{-1}$) and $-2.0\pm 1.3$~hPa~K$^{-1}$
(land $-2.2\pm 0.8$~hPa~K$^{-1}$, ocean $-1.9\pm 1.4$~hPa~K$^{-1}$) for the larger upper and lower panels, respectively.
}
\end{figure*}

In the equatorial land regions with high rainfall $--$ in the Amazon and Congo river basins, see point C in Fig.~\ref{figmap} $--$ the regressions were not
significant at $0.01$ probability level\footnote{On land, sea level pressure is not an empirically measured variable,
but is calculated from pressure $p_l(z_l)$, temperature $T_l(z_l)$ and the geopotential height $z_l$ of the land surface
assuming $\Gamma = 6.5$~K~km$^{-1}$ for $0 \le z \le z_l$, where $z = 0$ corresponds to the sea level.
This definition introduces a formal dependence of $p_{sl}$ (sea level pressure on land) on surface air temperature $T_l$, the strength
of which is directly proportional to $z_l$. That is, $p_{sl}$ diminishes with growing $T_l$ even if $p_l$ and, hence, the amount
of gas in the atmospheric column remains constant. Approximating the hydrostatic equation (\ref{he}) as $(p_l - p_{sl})/z_l = -p_l/h$, $h = RT_l/(Mg)$,
and taking the derivative of this equation over $T_l$  at constant $p_l$ we obtain $dp_{sl}/dT_l = (z_l/h) p_l/T_l$.
For the mean geopotential height $z_l = 0.6$~km of the tropical land, $p_l = 950$~hPa and $T_l = 295$~K we find
$dp_{sl}/dT_l = -0.2$~hPa~K$^{-1}$, i.e. about 20\% of the mean ratio established by us for the tropical land (Fig.~\ref{figmap}) is not related
to any air redistribution but is a formal consequence of the definition of $p_{sl}$.}. Where the regressions are significant, the largest (by absolute magnitude) regression slopes
tend to be concentrated in the regions of the largest SLP gradients, i.e. around the 15-20th latitudes (Fig.~\ref{figmap}).
These local dependences between $\widetilde{\Delta}p_s$ and $\widetilde{\Delta}T_s$ can be explained by the seasonal migration of the Hadley cells
where lower pressure is spatially associated with higher temperature (see Fig.~\ref{fighad}b below).
This explanation is supported by the fact that the tropical mean of the local
$\widetilde{\Delta}p_s/\widetilde{\Delta}T_s$ ratio, $-1.1$~hPa~K$^{-1}$ (Fig.~\ref{figmap}), is approximately equal to the
tropical mean ratio of the spatial differences $\Delta p_s/\Delta T_s = -1.3$~hPa~K$^{-1}$ (see Appendix~\ref{app}).
Likewise the result of \citet[][cf. their Figs.~2 and~8]{bayr13} $--$ that the long-term
trends in $p_s$ and $T_a$ have a similar ratio $-2.4$~hPa~K$^{-1}$ as their mean spatial differences $--$ indicates that these trends
reflect a shift in the form (e.g., widening) or displacement of the Hadley cells.

Outside the tropics where, in contrast to the tropics, areas of low pressure are at the same time areas of low temperature (particularly
the southern Ferrel cell), the seasonal relationship between pressure and temperature changes is generally less consistent than it is in the tropics
and somewhere it is reversed $--$ i.e., pressure and temperature rise or decline together (see point E in Fig.~\ref{figmap}).

Since the relationship between tropical pressure and temperature is apparently variable,
a model that assumes a constant ratio between temporal changes of pressure and temperature
cannot be used for predicting regional changes of pressure from changes in temperature in a warming or cooling climate.
Nor can a model based on a constant ratio between spatial differences of temperature and pressure successfully describe
the time-averaged circulation. Relative errors resulting from such models will be the largest where the pressure and
temperature variation are the smallest by absolute magnitude. \citet{lindzen87} emphasized how a distorted representation of the small pressure
gradients in the equatorial regions can mislead model-derived estimates of circulation and moisture convergence intensity.

\section{Vertical profiles of pressure differences}
\label{sdiff}

We will now discuss in a broader context the question of causality: is there a physical
mechanism by which differential heating at the surface could cause a surface pressure gradient?

\begin{figure*}[h]
\centerline{
\includegraphics[width=0.95\textwidth,angle=0,clip]{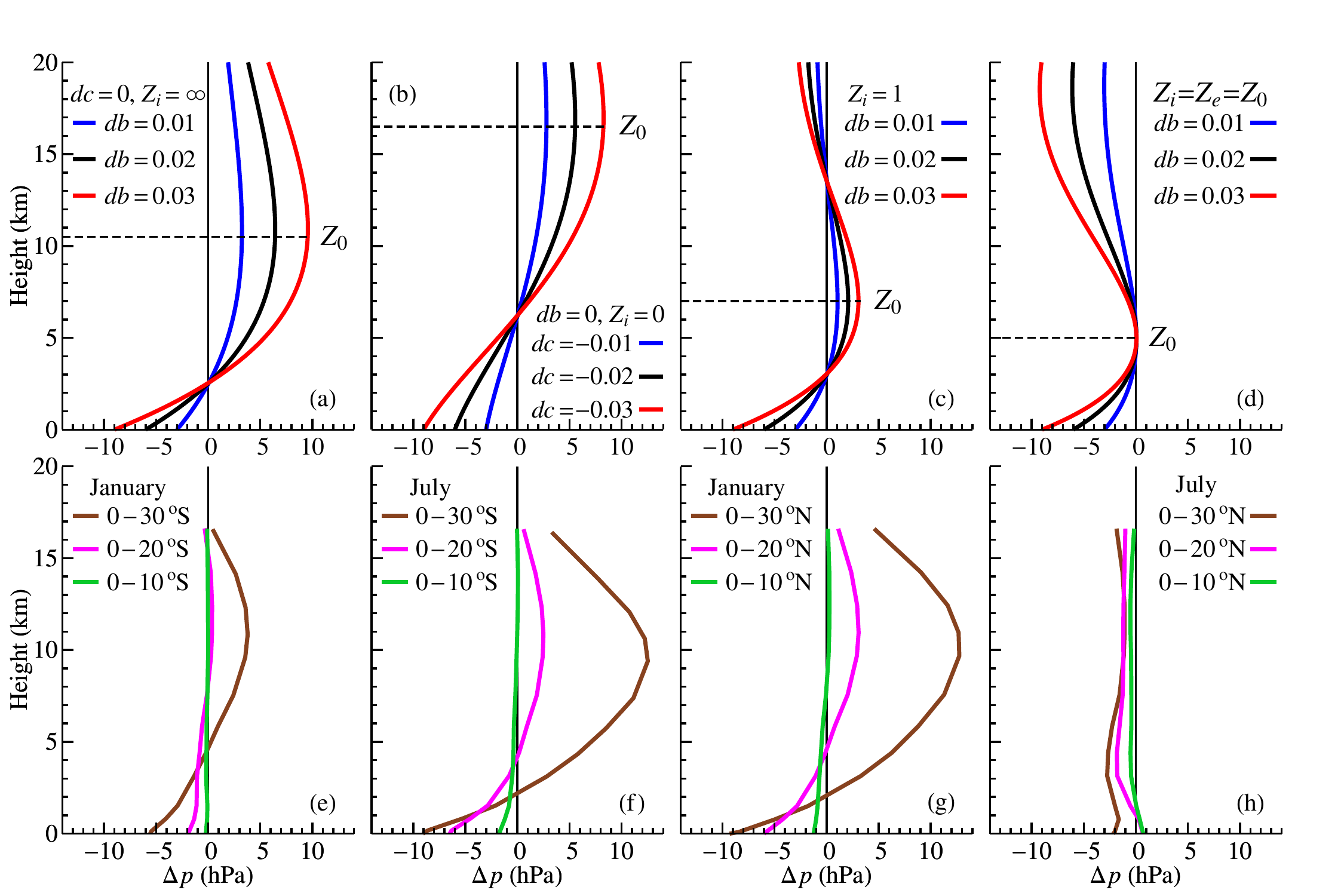}
}
\caption{\label{figdiff}
%\large
Vertical profiles of pressure differences $\Delta p(z)$ between air columns differing
in their lapse rate, surface pressure and temperature. Panels (a)-(d): theoretical
profiles (\ref{dp}) with $dp = \Delta p$, $da = \Delta p_s/p_s$,
$db = \Delta T_s/T_s$, $zdc = \Delta \Gamma/\Gamma_g$ (cf. \ref{diff}), $p_s = 1000$~hPa,
$T_s = 300$~K. In panels (a)-(d) $da = -0.003$, $-0.006$, $-0.009$ for the blue, black and red curves,
respectively. In each panel $da/db = constant$ for all the three curves. Dashed line $Z_0$ (\ref{Z0}) shows the height where the positive pressure difference in the
upper atmosphere is maximum, $\Delta p(Z_0) = \Delta p_0$ (\ref{dp0}). Note two isobaric heights in
panel (c). In panel (d) note that condition
$Z_i = Z_0$ (the atmosphere is horizontally isothermal where the positive pressure difference aloft is maximum)
yields $Z_i = Z_0 = Z_e = -2da/db = (-2da/dc)^{1/2}$, see (\ref{Ze}), (\ref{Z0}) and (\ref{Zi}),
and $\Delta p_0 = 0$, i.e. the pressure surplus aloft disappears.
Panels (e)-(h): real vertical profiles of zonally averaged pressure differences between the air columns at the equator and
the 10th, 20th and 30th latitudes in the Southern (e,f) and Northern (g,h) hemispheres
in January (e,g) and July (f,h).
E.g., the brown line in (e) shows the difference between the air column at the equator and at 30$\degree$S in January.
Note that while the theoretical curves (a-d) in each panel are chosen such that they have one and the same isobaric height $Z_e$ (i.e.,
they cross the line $\Delta p = 0$ at the same point), this varies for the real profiles (e-h).
}
\end{figure*}

In this section we will consider differentials in (\ref{diff}) as corresponding to small finite differences in respective variables ($\Delta p$,
$\Delta T$ and $\Delta \Gamma$) between two air columns that are separated along the meridian by a small finite distance $\Delta y$.
Then $dp = \Delta p(z)$ in (\ref{dp}) is a small pressure difference at a given height between the two air columns.
This difference has an extremum above the isobaric height $Z_e$ (\ref{Ze}) at a certain height $Z_0$ which is determined
by taking the derivative of (\ref{dp}) over $Z$ and equating it to zero, see (\ref{dp}), (\ref{Ze}) and (\ref{Zi}):
\beq\label{Z0}
\frac{\pt \Delta p}{\pt Z} = 0,\,\,\,da+Z_0db -\frac{1}{2}Z_0^2dc - db + Z_0dc =0,\,\,\,
Z_0 = 1+ Z_i \pm \sqrt{(Z_e-Z_i)^2+1}.
\eeq
At this height the pressure difference is equal to
\beq\label{dp0}
\Delta p_0 \equiv \Delta p(Z_0) = p_s e^{-Z_0} \left(da + Z_0db -\frac{1}{2}Z_0^2dc\right)=p_s e^{-Z_0} (db - Z_0 dc).
\eeq
Note that by definition when $\Delta p_0 = 0$ we have $Z_e = Z_0 = Z_i$.
As is clear from Fig.~\ref{figdiff},
where the vertical profiles of $\Delta p(z)$ (\ref{dp}) are shown for different
values of $da$, $db$ and $dc$, this extremum corresponds to the maximum pressure difference
between the air columns above the lower isobaric height.

When the vertical lapse rate is constant, $dc =0$, from (\ref{Z0}) we have $Z_0 = 1 -da/db$.
In this case, as is clear from (\ref{dp0}), for small values of $da/db \ll 1$ the magnitude of $\Delta p_0$ does not depend on $da$,
but is directly proportional to $db$, i.e. to $\Delta T_s$ (\ref{diff}) (Fig.~\ref{figdiff}a).
This means that under these particular conditions a surface temperature gradient directly determines the pressure gradient {\it in the upper atmosphere.}
In this sense there is no difference between surface temperature gradient and a gradient of lapse rate related to latent heat release
$--$ both can only determine a pressure surplus aloft, cf.~Fig.~\ref{figdiff}a,b and Eqs.~(\ref{a/b}) and (\ref{a/c}).
We emphasize that while the magnitude of the tropospheric pressure gradient
can be approximately specified from considerations of the hydrostatic balance and surface temperature gradients alone,
the magnitude of the surface pressure gradient cannot.

In the general case, the height of the extremum $Z_0$
as well as the ratio between the pressure
surplus aloft and the pressure shortage at the surface $\Delta p_0/\Delta p_s$
are functions of two parameters, the isobaric and isothermal heights $Z_i$ and $Z_e$.
Thus, when $Z_i$ and $Z_e$ are constant in space or time,
the ratio between the pressure surplus aloft and the pressure shortage at the surface in the warmer
column is constant as well:
the larger
the pressure surplus aloft, the larger the surface pressure shortage, with a direct proportionality between the two.
This is consistent with the
conventional thinking about differential heating, that the upper pressure surplus will cause air to diverge from
the warmer column, the total amount of gas will diminish and there appears a shortage of pressure
at the surface $\Delta p_s < 0$ .

\begin{figure*}[h]
\centerline{
\includegraphics[width=0.7\textwidth,angle=0,clip]{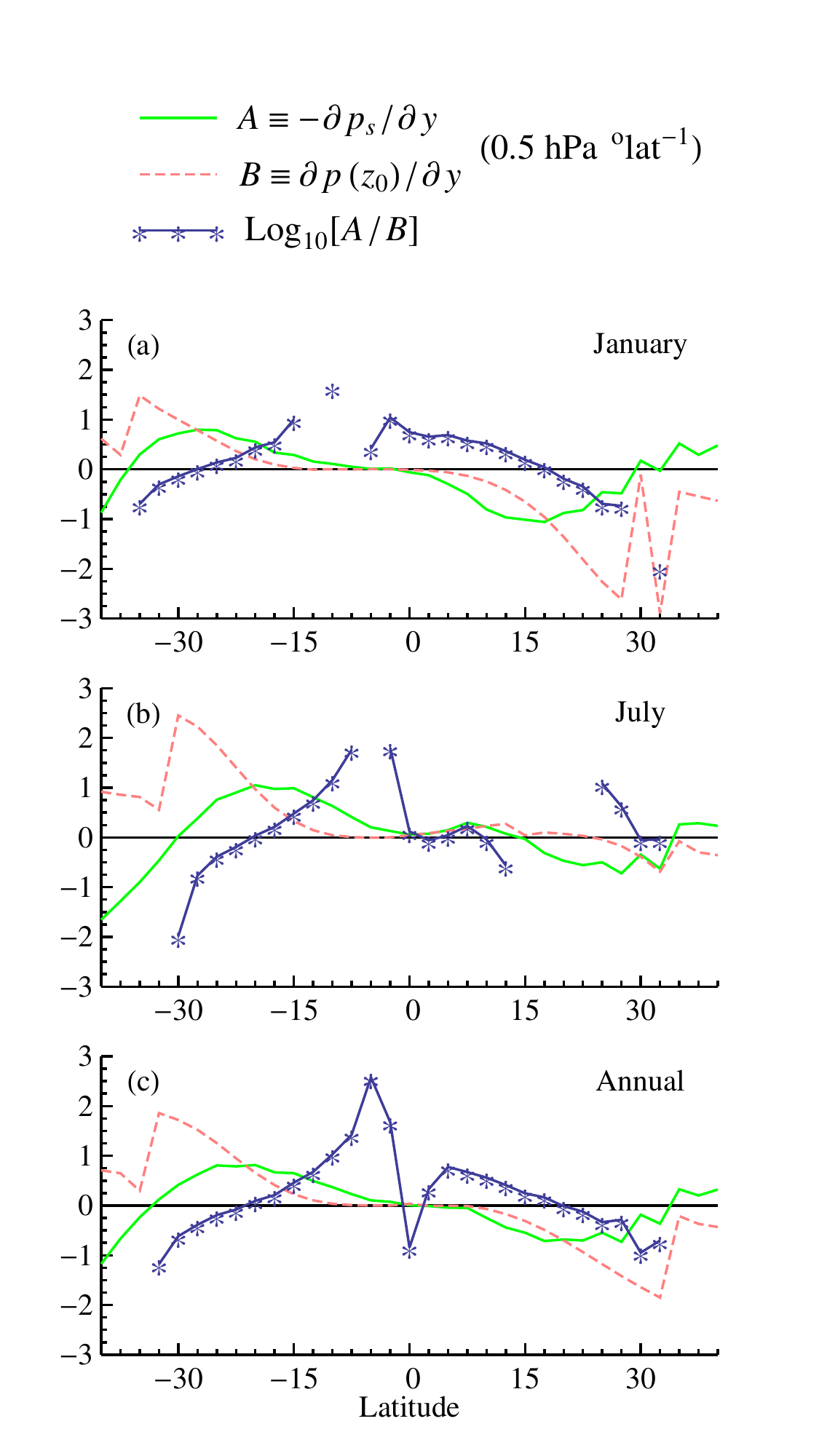}
}
\caption{\label{figdp0}
%\large
Logarithm of the ratio of the meridional gradient of SLP taken with the minus sign
$A \equiv -\pt p_s/\pt y$ to the meridional pressure gradient in the upper atmosphere $B \equiv \pt p(z_0)/\pt y$.
$B$ is calculated as the pressure difference $\Delta p_0$ at $z = z_0$ (\ref{dp0})
between two neighboring latitudes divided by 2.5$\degree$: $z_0$ is the height where $|\pt p(z)/\pt y - \pt p_s/\pt y|$ is maximum at a given $y$.
Missing values indicate latitudes where $A$ and $B$ are of different sign (and there is thus no isobaric height in the troposphere, cf. Fig.~\ref{figze}a,g).
}
\end{figure*}

This reasoning would be testable if it were
possible to specify $Z_i$ and $Z_e$ independently of the $da/db$ ratio. However, such an independent
specification apparently does not exist, while $Z_e$ varies significantly in space and time
(Figs.~\ref{figze}, \ref{figtemp}). The ratio between the surface pressure gradient and the maximum pressure gradient in the upper
atmosphere
\beq\label{p0}
\frac{\pt p_s/\pt y}{\pt p_0/\pt y} = -e^{Z_0}\frac{Z_e\left[1-Z_e/(2Z_i)\right]}{1-Z_0/Z_i}
\eeq
also varies within broad margins (Fig.~\ref{figdp0}). It is
larger at the equator than at the poleward ends of the cell: the larger the pressure gradient
aloft by absolute magnitude, the smaller, in relative terms, the pressure gradient at the surface (Fig.~\ref{figdp0}).

Our reading of current evidence and arguments is that a physical theory explaining how differential heating determines low-level pressure gradients does not exist.
That is to say, it remains impossible to link observed pressure gradients to gradients of air temperature using fundamental atmospheric constants and physical relationships.
Thus, any air circulation model attempting to reproduce low-level circulation based on differential heating physics must tune its key parameters (e.g., the $da/db$ ratio)
to fit with observations. Such a fitted model cannot readily be used to test the underlying relationships as their validity has already been assumed.
We propose that surface pressure and temperature gradients are generated primarily by water vapor dynamics.
We will now explain the physical mechanisms.

\section{Condensation-induced pressure differences}
\label{vapor}

The key physical proposition is that water vapor condensation in the moving air releases potential energy
at a rate $s$ (W~m$^{-3}$) proportional to air velocity in the direction of decreasing partial pressure of water vapor $p_v$
\citep{mg10,m13,mgn14}:
\beq\label{s}
s = -p \textbf{w}\nabla \gamma - \textbf{v}\nabla p_v,\,\,\,\gamma \equiv p_v/p.
\eeq
We consider zonally averaged stationary circulation where all variables depend on height $z$ and
distance along the meridian $y$;
$\textbf w$ and $\mathbf{v}$ are the vertical and horizontal (meridional) air velocities, respectively.
The first term describes condensation in the rising air. The second term describes
condensation or evaporation in the air moving along a horizontal temperature gradient.
Integrating $s$ over height in the entire atmosphere yields
precipitation $P$ per unit area of the Earth's surface in energy units $PRT$ (W~m$^{-2}$). With potential energy from condensation converted to the kinetic energy of atmospheric air,
$PRT$ should be equal to the independently estimated
total power of atmospheric circulation on Earth. This agrees well with observations \citep{m13,jas13}.

In hydrostatic equilibrium the kinetic power is generated by horizontal
pressure gradients only (the vertical pressure gradients are offset by the gravity force).
Integrating $s$ (\ref{s}) over the entire volume occupied
by the condensation-induced circulation we have
\beq\label{int}
-\int {\mathbf v}\nabla p dz dy = \int s dz dy = \int [-p \textbf{w}\nabla \gamma - {\mathbf v}\nabla p_v] dz dy.
\eeq
Eq.~(\ref{int}) formulates a constraint on the total kinetic power of a stationary circulation driven by condensation.
Our goal is now to show that under reasonable assumptions about the geometry of the circulation
and condensation areas Eq.~(\ref{int}) makes it possible to estimate surface pressure difference $\Delta p_s$
across the circulation.

\begin{figure*}[h]
\centerline{
\includegraphics[width=0.6\textwidth,angle=0,clip]{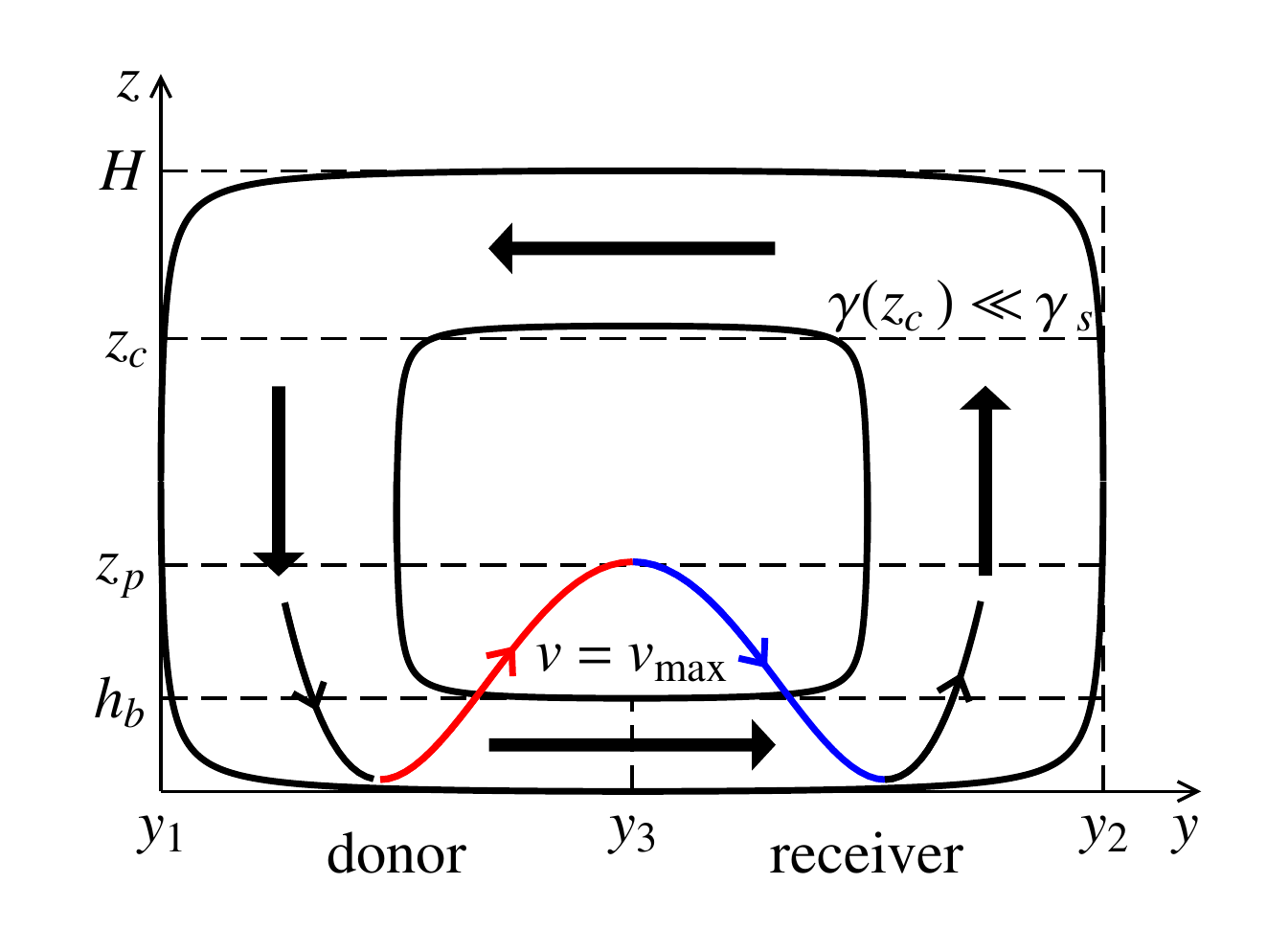}
}
\caption{\label{figc}
Donor and receiver areas. Thick arrows indicate large-scale air flows. Condensation occurs in the receiver area at $h_b \le z \le z_c$. Thin arrows
indicate turbulent eddies accompanying the large-scale flow with air parcels rising in the donor area with
a low (moist adiabatic) lapse rate (red curve) and descending in the receiver area with a higher (dry adiabatic) lapse rate (blue curve).
See text for other details.
}
\end{figure*}

At the outer borders of our circulation $y = y_1$ and $y = y_2$ meridional velocity $v$ is zero, $v(y_1) = v(y_2) = 0$ (Fig.~\ref{figc}).
Water vapor evaporated in the upstream part of the circulation
where the air descends is transported to the downstream part of the circulation where the air ascends and the imported water vapor condenses.
Reflecting this water vapor transport it is convenient to divide the circulation area into two parts, the donor and the receiver areas,
respectively. They are delimited by line $y = y_3$ where horizontal velocity $v$ is maximum $v(y_3) = v_{max}$ (Fig.~\ref{figc}).
For simplicity we assume the two parts to be of equal size. The length of the donor and receiver areas are respectively
$L_d = y_3 - y_1$ and $L_r = y_2 - y_3$, total length $L = 2 L_d =2L_r = y_2 - y_1$ (Fig.~\ref{figc}).

\begin{figure*}[h]
\centerline{
\includegraphics[width=0.6\textwidth,angle=0,clip]{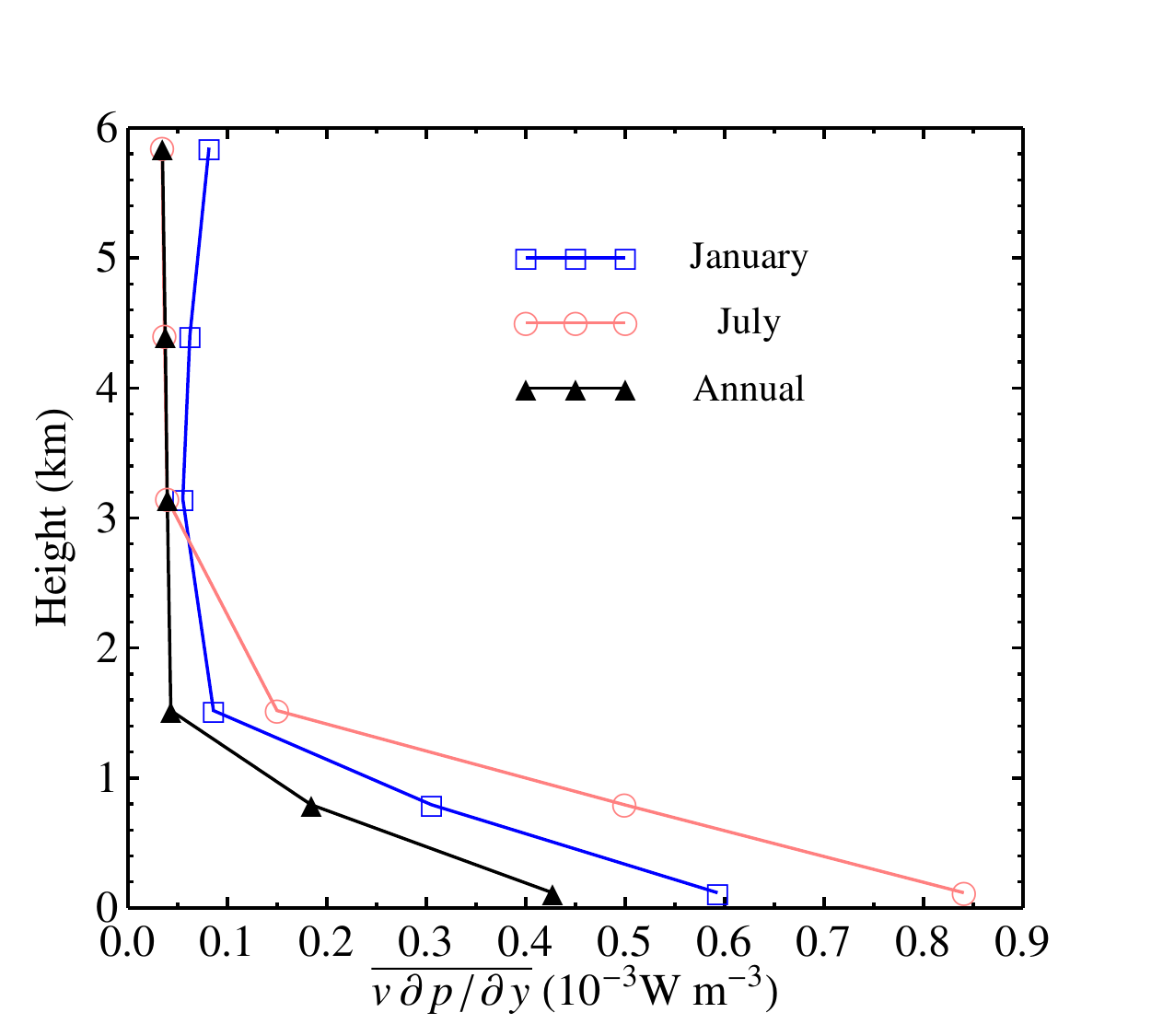}
}
\caption{\label{figvgp}
%\large
Tropical mean vertical profiles of the product of meridional velocity and meridional pressure gradient in the lower atmosphere
in January, July, and annually averaged. Profiles constructed by calculating
$v \pt p/\pt y$ in each gridpoint at different pressure levels as described and then averaging
from 30$\degree$S to 30$\degree$N. Monthly data from NCAR-NCEP reanalysis averaged for 1978-2013 (see Section~\ref{data} for more details).
}
\end{figure*}

We take into account that most part of kinetic energy is generated and dissipated in the narrow
layer near the surface $z \le h_b$ such that the first integral in (\ref{int})
can be written as
\beq\label{vgp}
\int {\mathbf v}\nabla p dz dy \equiv h_b \int (v_s \pt p_s/\pt y) dy,
\eeq
where $v_s$ and $p_s$ are velocity and pressure
at the surface and $h_b$ is the effective height of this layer. The product
$v \pt p/\pt y$ declines approximately linearly with increasing height up to 850 hPa (1.2~km) where it becomes
about one order of magnitude smaller than it is at the surface (Fig.~\ref{figvgp}). This means that $h_b \approx 0.6$~km in (\ref{vgp})
approximately coincides with the planetary boundary layer.
We also assume that the low-level air moves from the colder donor area to the warmer receiver area, such that
$\nabla p_v$ in (\ref{int}) describes the gradient of water vapor partial pressure owing to surface evaporation
that increases water vapor concentration in the surface layer $z \le z_s$ where the relative humidity
is less than unity, with $z_s$ being the saturation level. Since $z_s \approx h_b$ we can approximate
$\int {\mathbf v}\nabla p_v dz dy$ in (\ref{int}) by $h_b \int (v_s \pt p_{vs}/\pt y) dy$.
Using (\ref{vgp}) and a linear approximation $-\nabla p = \Delta p/L$, $\nabla p_v = \Delta p_v/L$ we can re-write (\ref{int}) as
\beq\label{int2}
(\Delta p + \Delta p_v) h_b \overline{v_s} = w_rL_r p\overline{\gamma_r},\,\,\,\overline{v_s}\equiv L^{-1}\int_{y_1}^{y_2} v dy,\,\,\,
w_r \equiv L_r^{-1}\int_{y_3}^{y_2}w(h_b)dy.
\eeq
Here $w_r$ is the mean vertical velocity at height $h_b$ and $\overline{\gamma_r} = (\gamma_2+\gamma_3)/2$ is the
mean relative partial pressure of water vapor at the surface in the receiver area, $\gamma_2$ and $\gamma_3$ are calculated,
respectively, at $y_2$ and $y_3$.
When deriving the right-hand part of (\ref{int2}) from (\ref{int}) we have taken into account that $pw$ is approximately constant
up to a height $z_c$ where most water vapor has condensed and $\gamma(z_c) \ll \gamma_s$\citep[see][their Eq.~A2]{jas13}.
We have also assumed that $s = - \textbf{v}\nabla p_v$ for $z \le h_b$ (no condensation below $h_b$).

On the other hand, from the integral continuity equation at the border of the donor and receiver areas we have
\beq\label{cont}
h_b v_{max} = w_r L_r,
\eeq
Assuming that $v$ increases approximately linearly from $v(y_1) = 0$ to $v(y_3)=v_{max}$ and then decreases linearly
to $v(y_2) = 0$ and neglecting the vertical variation in velocity in the boundary layer
$z \le h_b$ \citep{stevens02} we put $\overline{v_s}/v_{max} = 1/2$. Using this ratio, the expression for $\overline{\gamma_r}$ and (\ref{cont})
we are able to cancel velocities and linear scales in (\ref{int2}) to obtain
\beq\label{int3}
\Delta p_s = p_s (\gamma_1 + \gamma_3)=p_{v1}+p_{v3},\,\,\,z \le h_b.
\eeq
The drop of surface pressure in our condensation vortex is equal to the sum of water vapor partial pressures
at the two borders ($y_1$ and $y_3$) of the donor area (Fig.~\ref{figc}). Remarkably,
the pressure drop does not depend on the linear size $L$ of the circulation.

However, the main equation~(\ref{int}) from which (\ref{int3}) derives
does not indicate the
spatial scale of the air velocities and pressure gradients under consideration.
If the considered horizontal scale $L$ includes many condensation-induced vortices with chaotically
oriented pressure gradients, then the resulting large-scale pressure gradient and large-scale mean velocities $v$ and $w$ observed on scale $L$
will be zero.  How much of total potential energy released upon condensation is attributed to
kinetic energy generation on a particular linear scale is determined by the horizontal transfer of water vapor on
the considered scale. Indeed, if all moisture evaporated in one half of considered area is precipitated in the same half (without being transferred
to the second half), then the horizontal pressure gradient across the area will be zero. If, on the other hand, a certain part $R$ of
moisture evaporated in the donor area is exported to the receiver area, then the power of air circulation
generated on scale $L$ will be $K = R/C$ of total condensation power $C$ in the considered area.

From the mass balance equation for the water vapor we have
\beq
\label{mc}
P_d = E - R,\,\,\,P_r = E + R,
\eeq
where $P_d$ and $P_r$ are total precipitation in the donor and receiver areas (that we assumed to be of equal size),
$R$ is the amount of water vapor imported from the donor area to the receiver area and $E$ is evaporation
assumed to be the same in both areas. Horizontal transport of water vapor diminishes precipitation in the donor area and increases
it in the receiver area. Transfer coefficient $K$ can be retrieved from
the precipitation ratio $r$ between the two areas:
\beq\label{K}
K \equiv \frac{R}{P_r + P_d} = \frac{1}{2}\left(\frac{1 - r}{1+ r}\right),\,\,\,r \equiv \frac{P_d}{P_r}.
\eeq

The value of $K$ can be viewed as describing the proportion of time and space that the circulation in the considered area
takes the form shown in Fig.~\ref{figc}, while during the rest of time/space the horizontal pressure gradients on the considered area are zero.
With an account of the transfer coefficient our theoretical estimate for the surface pressure difference on a spatial scale $L$
becomes
\beq\label{da}
\Delta p_s = K p_s (\gamma_1 + \gamma_3) = \left(\frac{p_{v1}+p_{v3}}{2}\right) \left(\frac{1-r}{1+r}\right).
\eeq

\begin{figure*}[h]
\centerline{
\includegraphics[width=0.9\textwidth,angle=0,clip]{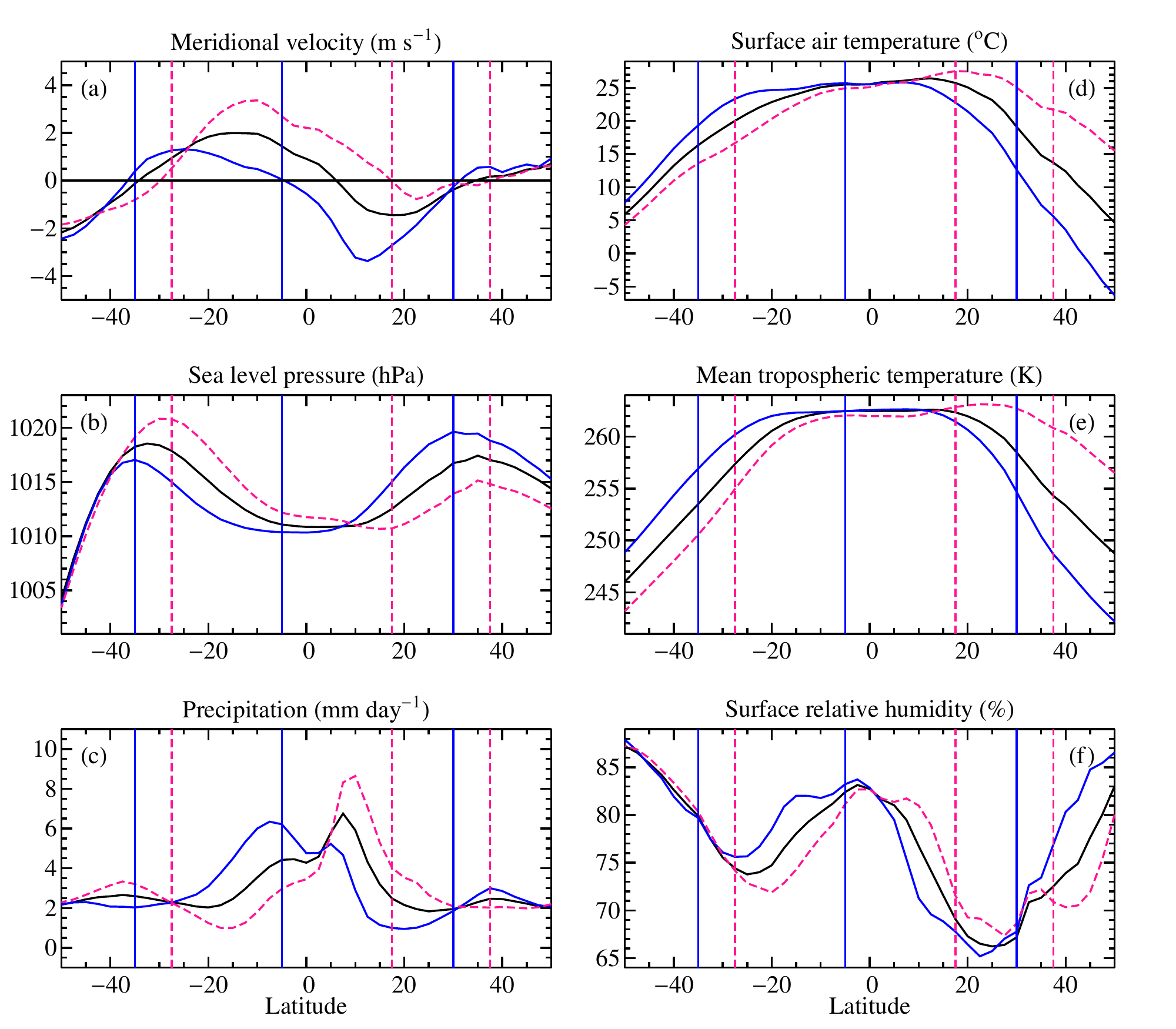}
}
\caption{\label{fighad}
%\large
Zonally averaged atmospheric parameters of Hadley cells. Solid black curve: annually averaged data,
solid blue curve: January, dashed pink curve: July. Vertical lines show the borders of the Southern
and Northern cells in January (solid blue) and July (dashed pink). These borders are defined as latitudes
where meridional velocity is zero (a). They simultaneously coincide with the two
poleward maxima (the outer borders) and the central minimum (the inner border) of SLP (b).
Monthly data from NCAR-NCEP reanalysis averaged for 1978-2013 (see Section~\ref{data} for more details).
}
\end{figure*}

We tested relationship (\ref{da}) with the zonally averaged data for the two Hadley cells (Fig.~\ref{fighad}).
For each month we computed the zonally averaged profile of SLP and meridional velocity $v$ at the surface.
For each month we defined the Northern and Southern cell as the areas where $v > 0$ and $v < 0$, respectively,
and computed the pressure difference $\Delta p_s$ between the poleward and the equatorial borders of each cell.
We calculated transfer coefficient $K$ using donor/receiver precipitation ratios as in (\ref{K}).
We calculated the mean surface water vapor partial pressure
$p_v = p\gamma$ by averaging the product of monthly mean relative humidity and
saturated water vapor partial pressure corresponding to the monthly mean surface air temperature in each grid point.
All estimated parameters are listed in Table~\ref{param}.

\begin{figure*}[h]
\centerline{
\includegraphics[width=0.9\textwidth,angle=0,clip]{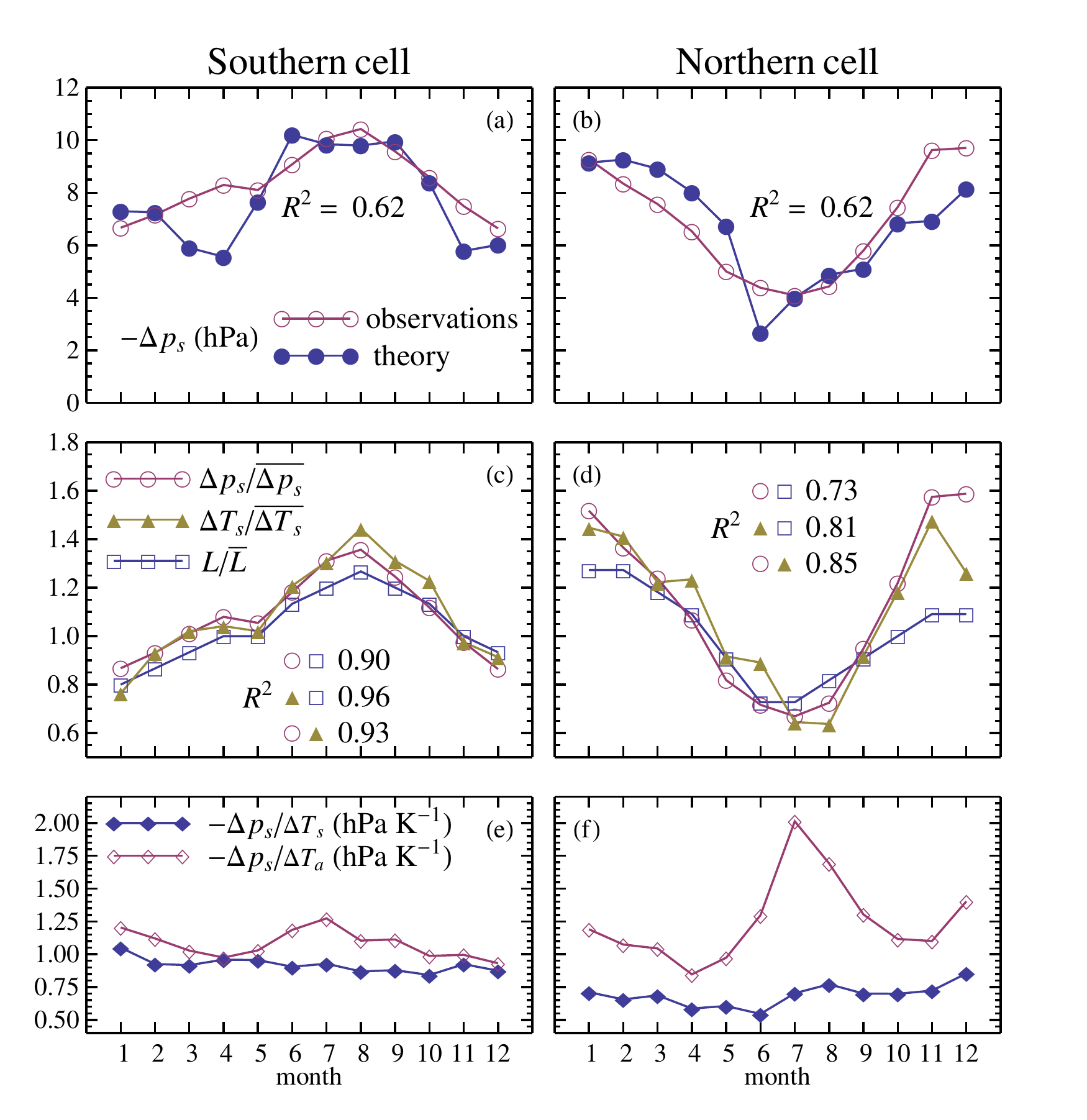}
}
\caption{\label{figdp}
%\large
Seasonal dynamics of pressure and temperature differences across the Southern (a,c,e) and Northern (b,d,f)
Hadley cells. a,b: observed and theoretically
estimated from Eq.~(\ref{da}) SLP differences $-\Delta p_s \equiv p_s(y_1) - p_s(y_2) \equiv p_{s1} - p_{s2}$
and the squared correlation coefficient for the ordinary least square regression between them.
c,d: Relative changes of SLP and temperature differences $\Delta p_s$ and $\Delta T_s$ and cell length
$L$ (all divided by their annual mean values denoted by overbar). Squared correlation coefficients
for the pairwise ordinary least square regressions between the variables are also shown.
e,f: Ratios of SLP difference to surface temperature $T_s$ and mean tropospheric temperature $T_a$ differences.
See Table~\ref{param} for all numerical values.
}
\end{figure*}

In Fig.~\ref{figdp} we plotted the observed monthly $\Delta p_s$ values versus the theoretical estimate (\ref{da})
for the Northern and Southern Hadley cells.
The annual mean theoretical estimates of $\Delta p_s$ are within 30\% of their observed values (Table~\ref{param}),
which can be considered a good agreement in the view of several simplifying assumptions that we have made.
The theoretical and empirical $\Delta p_s$ values display consistent changes throughout the year (Fig.~\ref{figdp}).
The order of magnitude of $\Delta p_s$ is set
by the partial pressure of water vapor $p_{vs}$ in the donor area, which changes little throughout the year.
The seasonal behavior of $\Delta p_s$ is governed by the transfer coefficient $K$, which
varies from $0.14$ to $0.27$ in the Southern cell and from $0.07$ to $0.29$ in the Northern cell.
It is higher during the colder season, when the cell is also larger
\citep[Fig.~\ref{figdp}c,d, see also][their Fig.~1]{dima03}.
While the poleward border $y_1$ moves towards the equator during the colder season, the near-equatorial
border $y_2$ spreads to the other hemisphere, such that the winter cell comprises a larger part of the precipitation peak
than the summer cell (Fig.~\ref{fighad}). This is manifested as an increase in the transfer coefficient $K$.

\section{Moisture transport and surface air temperature}
\label{temp}

The surface temperature differences $\Delta T_s$ associated with pressure differences $\Delta p_s$ are shown in Fig.~\ref{figdp}c,d.
In the tropical area, where the solar flux varies least with latitude compared to the extratropics,
the horizontal transport of moisture and, hence, latent heat should play a major role in the spatial distribution of temperature.
Water vapor evaporates in the donor area and condenses in the receiver area.
Thus the donor area exports, and the equatorial receiver area imports, significant amounts of energy in the form of latent
heat. How could this process influence surface temperature? We have a suggestion.

Consider an air parcel in the donor area that rises from the surface $z = 0$ up to a certain height $z_p$ (Fig.~\ref{figc}).
It starts from a surface pressure $p_{s1}$ and surface temperature $T_{s1}$, its temperature varies with a lapse rate $\Gamma_1$.
The parcel travels at this height towards the equator where it descends and returns to the surface
with a different lapse rate $\Gamma_2 > \Gamma_1$. A relevant example is the ascent with a moist adiabatic and descent with a dry adiabatic
lapse rate. Upon the descent, at the surface this parcel will have a higher temperature $T_{s2} = T_{s1} + \Delta T_s > T_{s1}$.
It will also have a lower pressure $p_{s2} = p_{s1}+\Delta p_s < p_{s1}$. This is because in the descending parcel
being on average warmer than the ascending parcel, pressure grows with diminishing height more slowly (its pressure
scale height $h$ (\ref{he}) is larger). Therefore, for such
a process to be possible, the area where the warmer parcel descends must have a lower surface pressure than where it started its ascent.
(A remarkable example of such descending motion in a warm low pressure area occurs in the eyes of
the tropical storms, which are both warmer than the zone of intense convection at the windwall
\citep{montgomery06} and have lower pressure \citep{mg11}.)

If pressure and temperature vary considerably less at $z_p$ (the height at which the parcel moves) than at the surface,
this height can be considered as both isobaric and isothermal, $z_p = z_e = z_i$.
This condition allows one to find the difference in the parcel's temperatures at the surface
from the known values of $\Delta p_s$ and $\Delta \Gamma$ using (\ref{Zi}) and (\ref{a/b}):
\beq\label{par}
Z_p = Z_e = Z_i = \frac{db}{dc}=-2\frac{da}{db},\,\,\,
db = \sqrt{-2da \, dc},\,\,\,\Delta T_s = T_s\sqrt{-2\frac{\Delta p_s}{p_s}\frac{\Delta \Gamma}{\Gamma_g}}.
\eeq
Taking $\Delta p_s = -7.7$~hPa for the mean difference between the inner and outer ends of the Hadley cell (Table~\ref{param})
and $\Delta \Gamma = 4$~K~km$^{-1}$ equal to the difference between the dry adiabatic lapse rate
and moist adiabatic lapse rate at $T \approx 283$~K \citep[][their~Fig.~4e]{mg10} with $p_s = 1013$~hPa and $T_s = 298$~K we obtain
from (\ref{par}) $\Delta T_s = 12.6$~K, ratio $\Delta p_s/\Delta T_s = -0.6$~hPa~K$^{-1}$ and height $z_p \equiv Z_p h_s = 3.2$~km.

Theoretical estimate of the ratio $\Delta p_s/\Delta T_s = -0.6$~hPa~K$^{-1}$  is smaller by absolute magnitude than the observed
(the mean annual ratio is $-0.9$~hPa~K$^{-1}$ for the Southern and $-0.7$~hPa~K$^{-1}$ for the Northern cell), while
the estimated temperature difference $12.6$~K is larger than the observed ($\Delta T_s = 8.3$~K for the Southern and $9.0$~K for the Northern cell).
Another discrepancy between the theory and observations is that the vertical mixing apparently spreads the temperature difference
well above the parcel height $z_p$. While there is indeed a local minimum of the temperature difference between the equator and the wider tropics
$z_p \approx 3$~km (Fig.~\ref{figtemp}a-e), this height is not strictly isothermal.

\begin{figure*}[h]
\centerline{
\includegraphics[width=0.60\textwidth,angle=0,clip]{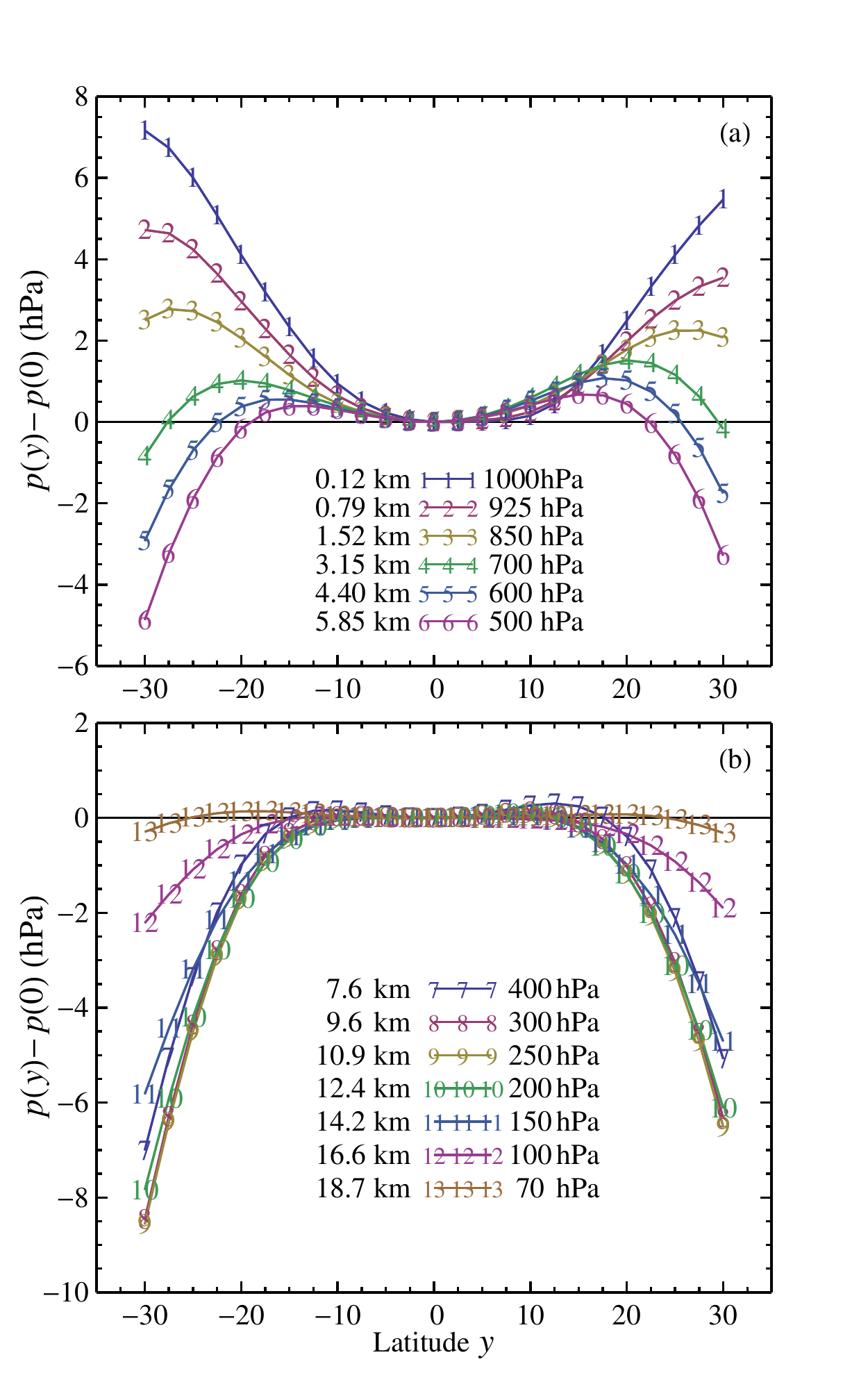}
}
\caption{\label{figlevp}
%\large
Annual mean difference $p(y) - p(0)$ between pressure at latitude $y$
and the equator at different heights. E.g. curve 1 in (a) shows meridional pressure
variation at height $z = 0.12$~km, which is equal to the tropical mean geopotential height
of pressure level 1000 hPa. Note that the pressure difference between the 30th latitudes
and the equator approaches zero for $z = 3.15$~km (pressure level 700 hPa).
Monthly data from NCAR-NCEP reanalysis averaged for 1978-2013 (see Section~\ref{data} for more details).
}
\end{figure*}

On the other hand, theoretical result (\ref{par}) agrees with the observations in several essential ways.
First, the pressure difference between the 30th latitudes and the equator
at the estimated height $z_p = 3.2$~km is close to zero and indeed much smaller than at the surface
(Fig.~\ref{figlevp}), supporting our assumption that for the considered air parcel $z_p = z_e$. Note that for Eq.~(\ref{par}) to hold, we do not need
$z_p$ to be the local isobaric height at any point $--$ we have only demanded that the pressure difference
along $z_p$ between the areas where the parcel ascends and where it descends is negligible compared
to the pressure difference at the surface.
Second, the atmospheric layer up to $z_p = 3.2$~km does indeed
represent the layer where the lapse rate increases from the wider tropics to the equator (Fig.~\ref{figlap}). It is in this low layer
that the equator, despite being the hotspot of rainfall and convection, has a steeper lapse rate than the rest of the tropics.
The equatorial lapse rate becomes moist adiabatic only starting from about 5~km \citep{mapes01}.
The descending motion of the low-level air parcels transporting latent heat
from the donor area provides an explanation to this remarkable feature.
Third, height of about 3~km represents the upper boundary of the trade wind inversion layer \citep{schub95}. Shallow convective clouds forming in this layer
represent a prominent feature of tropical convection $--$ in fact they are one of the three dominant convective modes \citep{johnson99}.
This shallow convection is more common to the poleward ends of the Hadley cells and is absent near the equator supporting the idea that the ascending
motion driving the low-level convection is concentrated in that area. Thus, the existence of moist air parcels rising around the 30th latititudes
up to 3~km and descending much closer to the equator does not contradict what we know about the tropical cloud cover.
Horizontal transport of latent heat and its conversion to sensible heat in the lower atmosphere near the equator thus appears able
to explain the observed surface temperature distributions.

\section{Discussion}

In the literature surface pressure gradients are discussed as {\it determined} or {\it generated}
by gradients in sea surface temperature \citep{lindzen87,sobel06,an11}.
For example, \citet[][p.~324]{sobel06} in their discussion of the model of \citet{lindzen87} noted that surface temperature
determines temperature in the atmospheric boundary layer, which, in its turn, determines
surface pressure via a hydrostatic relationship. Likewise \citet{bayr13} characterized the differential heating of the planetary surface
as a driver of changes in surface pressure.

Here we have revisited the concept of differential heating in a hydrostatic atmosphere.
As considered in Section~\ref{sdiff}, under certain conditions the surface temperature gradients can
indeed approximately determine pressure gradients, but only in the upper atmosphere.
However, the magnitude of the surface pressure gradient cannot be deduced from the magnitude of surface temperature gradient
unless some additional postulates are made that would a priori specify a relationship between the two.
In particular, a linear relationship between surface gradients of temperature and pressure is contingent upon the existence of
a constant isobaric height where pressure does not vary. The existence of such a constant height was postulated
in the models of \citet{lindzen87} and \citet{bayr13}.

Here we have used empirical evidence to demonstrate that there is neither a constant isobaric height in the tropics
nor is such a constancy a reasonable zero-order approximation.
The isobaric height (defined as the height where the meridional pressure gradient is zero) varies from zero at the 30th latitudes to over 16~km near the equator. Its magnitude cannot
be deduced from any fundamental atmospheric parameters $--$ the physical model of \citet{bayr13} proposing the height of the troposphere
as a universal isobaric height was not correct (Section~\ref{bayr}).
We showed that the observed ratio between surface pressure and temperature gradients defines the magnitude
of the isobaric height if one more essential parameter, the isothermal height, is known (Eq.~\ref{a/b}).
We thus conclude that the existence of a relationship between surface pressure and temperature (with warm air having low pressure)
is not an argument that surface pressure gradients are driven by differential heating.
Conversely, the concept of differential heating based on a constant isobaric height cannot explain
why the surface temperature and pressure gradients across the tropics have the magnitudes observed.

In contrast, we have demonstrated that evaporation and condensation
can produce the observed SLP differences of the order of $\Delta p_s \sim 10$~hPa in the zonally averaged Hadley cells (Section \ref{vapor}).
The scale of the pressure differences is set by the mean partial pressure
of the water vapor in the donor area. Their actual magnitude depends on the efficiency $K$ of horizontal moisture transport (\ref{da}).
Coefficient $K$ (\ref{K}) describes the ratio of the intensity of condensation associated with horizontal moisture
transport on a length scale comparable to the length of the Hadley cell to the intensity of condensation associated with smaller-scale local eddies.
The efficiency $K$ of horizontal moisture transport grows with the increasing linear size of the Hadley cell.
In winter cells both in Northern and Southern hemispheres $K$ reaches its maximum value of $0.3$. In the smaller summer cells
$K$ falls to about $0.1$ (Table~\ref{param}). The maximum possible value of $K = 1/2$ would imply that all water vapor evaporated in the poleward
half of the Hadley cell (the donor part) has been transported to the equatorial counterpart and precipitated there.
$K = 0$ means that all evaporated moisture precipitates locally -- i.e. that the characteristic transport length is much less than
the cell length $L$. In such a case, when condensation is spatially uniform, the vapor sink obviously does not produce any large-scale
pressure gradient.

What determines the seasonal changes in $K$? Condensation in the rising air must, by mass conservation constraints, always involve some horizontal
air motion. If we have an isothermal
surface uniformly heated by the Sun convection can occur just by symmetry breaking: if the air begins
to condense in one place, there will be rising motion and horizontal import of moisture to the area of condensation.
Several studies, most importantly \citet{holloway10} for an equatorial island and \citet{sharkov12} in the context of
tropical cyclones linked the probability of convective rain
to the amount of water vapor in the atmospheric column. The higher the amount of water vapor, the higher the probability
of (intense) convection.
Any small differences in solar radiation over an otherwise uniform oceanic surface
will translate into differences in the accumulated flux of evaporated water vapor. Since likelihood of rain
rises sharply with columnar water vapor content \citep[e.g.,][their Fig.~10b]{holloway10}, the area receiving more solar flux will develop
convection sooner than the area that receives less. This will lead to a drop of pressure and a horizontal transport of moisture towards
the area where condensation takes place \citep[see also discussion by][]{jhm14}. The pressure gradient generated through this process enhances horizontal
motion and moisture transport which reinforces and enhances the pressure gradient itself. Therefore even a small
gradient in solar radiation can in principle cause significant spatial gradients in condensation intensity. In such a case, condensation
will be more spatially uniform (i.e., $K$ will be lower) in summer than in winter cells, in agreement with observations (Table \ref{param}).

With condensation intensity depending on minor differences in local water vapor amounts,
natural forests with their intense evapotranspiration can play a much larger role in determining the position
of active convective zones than is generally recognized \citep{mg07,jhm14}. For example, the on-going discussion
concerning possible slow-down of the Walker circulation focuses on the relationships
between sea level temperature and pressure \citep[e.g.,][]{tokinaga12}, while the concurrent large-scale deforestation on the Maritime
Continent and the associated changes in evapotranspiration are never considered as possible drivers of
the regional changes in convection.

We have additionally suggested that horizontal transport of latent heat from the outskirts of the Hadley cells
(donor areas) toward their inner equatorial parts (receiver areas)
can lead to formation of a horizontal surface temperature gradient (Section~\ref{temp}). Latent heat captured as
water vapor at the 30th latitudes is transported by the converging air towards the equator. Convective eddies where the air descends
dry adiabatically ensure that part of this heat is returned to the surface in sensible form.
This process may partially account for the fact that the equator in the lower atmosphere (up to 850 hPa) has a steeper
lapse rate than the rest of the tropics (Fig.~\ref{figlap}).

In the extratropics latent heat release was discussed as a mechanism stabilising
the upper tropospheric temperatures during winter time in the extratropics \citep{herman08}.
In the tropics, the effects of evaporation and latent heat release have been considered extensively in the context of climate stability
\citep{wallace92,rama91,bates99,caballero01,bates12}. \citet{wallace92} observed that evaporation can cool the surface as the latent
heat released in the upper atmosphere will be rapidly mixed in the horizontal dimension cooling the warm surface
more than it warms cool surfaces. However, this cooling mechanism considers only {\it export} of local latent heat resulting
 from condensation of moisture evaporated in the region of ascent.
Meanwhile, condensation in the ascending air is necessarily accompanied by {\it import} of moisture and, hence, latent heat
from the adjacent areas to the area of ascent. If, as we proposed, this additional latent heat is released and converted
to sensible heat in low-level eddies in the zone of convection, the outcome
may be not a uniform temperature distribution but, rather, a creation of a surface temperature gradient. We have shown
that moist air parcels in convective eddies rising to the height of the trade wind inversion and descending in the low pressure equatorial area
can produce temperature gradients of magnitudes close to the observed.

Since higher temperatures are associated with higher atmospheric content of water vapor, a surface temperature gradient
 can be another mechanism responsible for the spatial non-uniformity of condensation intensity besides the surface gradient in
absorbed solar radiation. If horizontal transport of latent heat is a major factor determining the surface temperature gradients in the tropics,
this can provide an alternative explanation for the relative constancy of near equatorial temperatures \citep{wallace92}.
Suppose the extratropics cooled compared to the equator. This enhanced the temperature difference
between the equator and the tropics and led to an extra import of latent heat towards the equator. In the result,
in the new cooler climate the equator cooled less than the tropics because of this extra heat.
Conversely if the extratropics warm, this leads to a decline in latent heat transport towards the equator,
such that in the new warmer climate the equator warms less. In the result equatorial temperatures become more stable
than at higher latitudes with respect to temperature fluctuations originating in the extratropics.
In summary, we believe that the perspectives opened by the concept of condensation-driven winds
merit further investigations.

\appendix
\nonumber
\section{Appendix: Relationship between $T_s$ and $T_a$}
\label{app}
%\large
The relationship between surface temperature $T_s$ and the mean temperature $T_a(Z)$ of the atmospheric column below $Z$
can be derived from (\ref{T}) and the hydrostatic equation (\ref{he}):
\beq\label{Ta}
T_a(Z) \equiv \frac{\int_0^{Z} T(Z) \rho dZ}{\int_0^{Z} \rho dZ} = \frac{T_s}{1+c}\frac{1-e^{-cZ}e^{-Z}e^{-cZ^2/2}}{1-e^{-{Z}}e^{-c {Z}^2/2}},
\,\,\,Z \equiv \frac{z}{h_s},\,\,\,cZ \ll 1.
\eeq
Expanding (\ref{Ta}) over $c$ and keeping the linear term we have
\beq\label{Ta2}
T_a = T_s\left[1 - c \left(1 - \frac{Z}{e^Z-1}\right)\right].
\eeq

\begin{figure*}[h]
\begin{minipage}[h]{0.49\textwidth}
\centerline{
\includegraphics[width=0.99\textwidth,angle=0,clip]{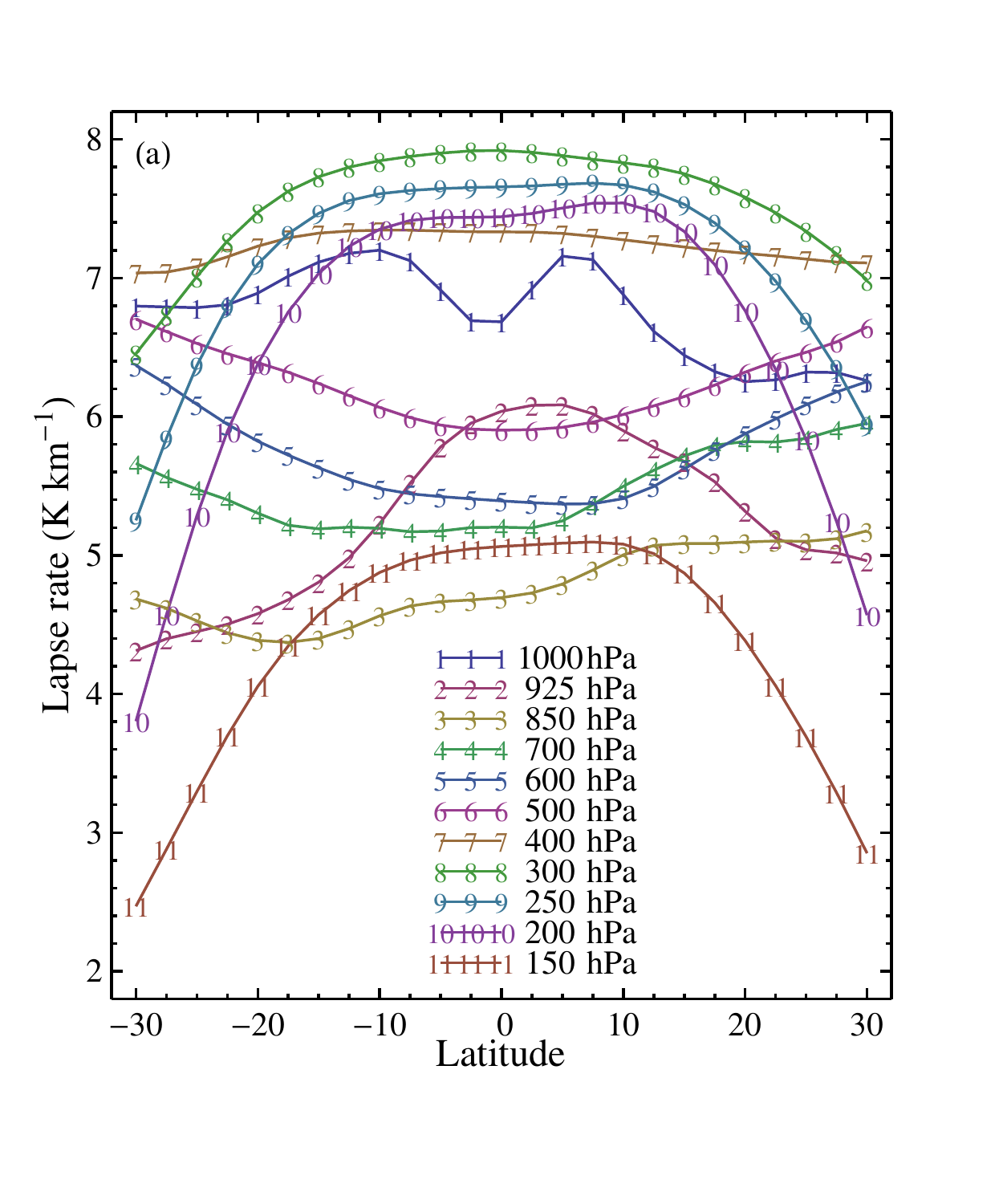}
}
\end{minipage}
\begin{minipage}[h]{0.49\textwidth}
\centerline{
\includegraphics[width=0.99\textwidth,angle=0,clip]{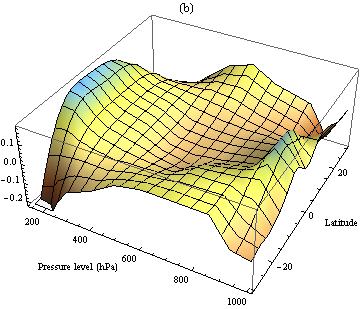}
}
\end{minipage}
\caption{\label{figlap}
%\large
Annual mean latitudinal profiles of the air temperature lapse rate on different
pressure levels. For example, curve 1 in (a) shows the mean lapse rate between 1000 hPa
and 925 hPa; curve 2 $--$ between 925 hPa and 850 hPa; curve 11 $--$ between 150 and 100 hPa.
The tropical mean lapse rate (the temperature difference between 1000 hPa
and 100 hPa levels divided by the difference in the geopotential heights and averaged
from 30$\degree$S to 30$\degree$N) is 6.0~K~km$^{-1}$. Panel (b) shows the relative variation -- at each pressure level the lapse rate
at a given latitude is divided by the mean lapse rate at this level (averaged between
30$\degree$S and 30$\degree$N). The equator has a higher lapse rate than the 30th latitudes in the lower and upper
-- but not the middle -- troposphere.
}
\end{figure*}

Taking the derivative of (\ref{Ta2}) over $T_s$ and $c$ we obtain:
\beq\label{c}
dc = \frac{db-dn}{1 - Z/(e^Z-1)},\,\,\,dn \equiv \frac{dT_a}{T_a}.
\eeq
For the height of the tropical troposphere $z = H = 16.5$~km, $T_s = 298$~K and $\Gamma = 6.0$~K~km$^{-1}$ (Fig.~\ref{figlap})
we have $Z=1.9$, $c = 0.18$, $1 - Z/(e^Z-1) = 0.66$ and obtain from
(\ref{Ta2}) and (\ref{c})
\beq\label{Ta3}
T_a = 0.88 T_s,\,\,\,db = dn + 0.66 dc,\,\,\,\frac{dT_s}{dT_a} = \frac{1}{0.88}\left(1+0.66 \frac{d\Gamma}{dT_a}\frac{T_a}{\Gamma_g}\right).
\eeq
The mean tropospheric $\overline{T_a} = 262$~K in the tropics estimated from (\ref{Ta3}) agrees
with the tropical mean $\overline{T_a} = 261$~K that we estimate from the TTT data of
\citet{mears09} and with $\overline{T_a} = 263.6$~K cited by \citet{bayr13}.

From (\ref{Ta3}) we can see that the relative changes $db$ and $dn$ of $T_s$ and $T_a$ coincide,
$db = dn$, if only the lapse rate does not vary, $dc = 0$. In the tropical atmosphere
this is not the case: areas with higher $T_s$ ($db > 0$) have a higher lapse rate $dc > 0$ (Fig.~\ref{figlap}).
Therefore, in the tropics $db > dn$. Table~\ref{tab2} lists the results of the reduced major axis regression
of $\Delta T_s = T_s - \overline{T_s}$ on $\Delta T_a = T_a - \overline{T_a}$ in the tropics (from 27.5$\degree$S to 27.5$\degree$N)
for different months on land and in the ocean. On average we have $\Delta T_s/\Delta T_a = 1.8$.
Therefore, the result of \citet{bayr13} $\Delta p_s/\Delta T_a = -2.4$~hPa~K$^{-1}$ corresponds to
$\Delta p_s/\Delta T_s = (\Delta p_s/\Delta T_a)/(\Delta T_s/\Delta T_a) = -1.3$~hPa~K$^{-1}$
as considered in Section~\ref{data}.

\newpage
\section{Tables}
\begin{table}[!h]\tiny
\caption{\label{param}Parameters of Eq.~(\ref{da}) and other relevant parameters of Hadley cells: $y_1$ and $y_2$ are the outer borders of the donor and receiver areas (Fig.~\ref{figc}), respectively;
$L$ cell length, $P$, $P_d$ and $P_r$ are mean precipitation in the cell as a whole, in the donor and receiver areas, respectively;
$K$ is the moisture transport coefficient (\ref{K}); $p_{v1}$ and $p_{v3}$ are partial pressures of water vapor at the surface at $y_1$
and $y_3 \equiv (y_2 + y_1)/2$ (the inner border of the donor and receiver areas); $p_{s1}$,	$T_{s1}$, 	$T_{a1}$ are
surface pressure, surface air temperature and mean tropospheric temperature, respectively, at $y_1$;
$\Delta p_s$, $\Delta T_s$ and	$\Delta T_a$ are differences in respective variables at $y_1$ and $y_2$ (e.g., $\Delta p_s = p_{s2} - p_{s1}$);
	$\Delta p_s*$ is the theoretical estimate (\ref{da}) of $\Delta p_s$.
}
\begin{center}
\begin{tabular}{lrrrrrrrrrrrrrrrr}
\toprule
Time &	$y_1$ &	$y_2$ &	$L$ &	$P$ &	$P_d$ &	$P_a$ &	$K$ &	$p_{v1}$ &	$p_{v3}$ &	$p_{s1}$ &	$T_{s1}$ &	$T_{a1}$ &	$\Delta p_s o$ &	$\Delta p_s t$ &	$\Delta T_s$ &	$\Delta T_a$         \\
\cmidrule(r){2-4}\cmidrule(r){5-7}\cmidrule(r){9-11}\cmidrule(r){12-13}\cmidrule(r){14-15}\cmidrule(r){16-17}
&\multicolumn{3}{c}{$\degree$lat}& \multicolumn{3}{c}{mm day$^{-1}$}& &\multicolumn{3}{c}{hPa}	 &	\multicolumn{2}{c}{K}&\multicolumn{2}{c}{hPa}& \multicolumn{2}{c}{K}\\
\midrule\multicolumn{17}{c}{\it Southern cell}\\
Jan &	$-35.0$ &	 -5.0 &	 30.0 &	 3.8 &	 2.4 &	 5.0 &	 0.17 &	 17.4 &	 24.0 &	 1017.0 &	 292.3 &	 256.9 &	$-6.67$ &	$-7.18$ &	  6.3 &	   5.5                                                        \\
Feb &	$-37.5$ &	 -5.0 &	 32.5 &	 3.9 &	 2.6 &	 5.3 &	 0.18 &	 16.5 &	 24.7 &	 1017.5 &	 291.2 &	 256.1 &	$-7.16$ &	$-7.26$ &	  7.7 &	   6.4                                                        \\
Mar &	$-37.5$ &	 -2.5 &	 35.0 &	 3.9 &	 2.6 &	 4.8 &	 0.15 &	 15.9 &	 24.2 &	 1018.0 &	 290.6 &	 255.1 &	$-7.77$ &	$-5.90$ &	  8.5 &	   7.6                                                        \\
Apr &	$-35.0$ &	  2.5 &	 37.5 &	 3.8 &	 2.6 &	 4.6 &	 0.14 &	 15.9 &	 24.0 &	 1018.5 &	 290.8 &	 254.4 &	$-8.30$ &	$-5.60$ &	  8.7 &	   8.5                                                        \\
May &	$-30.0$ &	  7.5 &	 37.5 &	 3.7 &	 2.2 &	 4.9 &	 0.19 &	 15.8 &	 24.9 &	 1018.7 &	 291.1 &	 255.0 &	$-8.10$ &	$-7.63$ &	  8.5 &	   7.9                                                        \\
Jun &	$-27.5$ &	 15.0 &	 42.5 &	 3.6 &	 1.9 &	 5.4 &	 0.24 &	 15.2 &	 26.4 &	 1019.8 &	 290.5 &	 255.3 &	$-9.08$ &	$-10.1$0 &	 10.0 &	   7.7                                                       \\
Jul &	$-27.5$ &	 17.5 &	 45.0 &	 3.6 &	 1.7 &	 5.2 &	 0.25 &	 14.2 &	 25.3 &	 1020.8 &	 289.7 &	 254.9 &	$-10.1$0 &	$-9.89$ &	 10.9 &	   7.9                                                       \\
Aug &	$-30.0$ &	 17.5 &	 47.5 &	 3.4 &	 1.4 &	 4.9 &	 0.27 &	 13.1 &	 24.0 &	 1020.9 &	 288.6 &	 253.5 &	$-10.4$0 &	$-9.87$ &	 12.0 &	   9.4                                                       \\
Sep &	$-30.0$ &	 15.0 &	 45.0 &	 3.3 &	 1.5 &	 4.9 &	 0.26 &	 13.4 &	 24.5 &	 1020.3 &	 289.3 &	 254.0 &	$-9.57$ &	$-9.97$ &	 10.9 &	   8.6                                                        \\
Oct &	$-32.5$ &	 10.0 &	 42.5 &	 3.2 &	 1.8 &	 4.5 &	 0.22 &	 13.6 &	 24.6 &	 1019.5 &	 289.1 &	 253.6 &	$-8.59$ &	$-8.34$ &	 10.2 &	   8.7                                                        \\
Nov &	$-32.5$ &	  5.0 &	 37.5 &	 3.1 &	 2.1 &	 3.9 &	 0.15 &	 15.0 &	 23.4 &	 1018.0 &	 290.7 &	 255.0 &	$-7.49$ &	$-5.70$ &	  8.1 &	   7.5                                                        \\
Dec &	$-35.0$ &	  0.0 &	 35.0 &	 3.5 &	 2.4 &	 4.4 &	 0.15 &	 15.8 &	 23.8 &	 1017.0 &	 290.8 &	 255.4 &	$-6.64$ &	$-6.02$ &	  7.6 &	   7.1                                                        \\
Ann &	$-32.5$ &	  5.0 &	 37.5 &	 3.3 &	 2.2 &	 4.0 &	 0.14 &	 15.3 &	 23.2 &	 1018.5 &	 290.7 &	 254.8 &	$-7.69$ &	$-5.38$ &	  8.3 &	   7.7                                                        \\
\multicolumn{17}{c}{\it Northern cell}\\
Jan &	 30.0 &	 -5.0 &	 35.0 &	 3.0 &	 1.3 &	 4.5 &	 0.28 &	 11.0 &	 21.9 &	 1019.6 &	 285.6 &	 254.7 &	$-9.28$ &	$-9.23$ &	 13.0 &	   7.8                                                         \\
Feb &	 30.0 &	 -5.0 &	 35.0 &	 2.8 &	 1.1 &	 4.1 &	 0.29 &	 11.0 &	 21.7 &	 1018.6 &	 286.2 &	 254.7 &	$-8.35$ &	$-9.34$ &	 12.7 &	   7.8                                                         \\
Mar &	 30.0 &	 -2.5 &	 32.5 &	 2.8 &	 1.1 &	 3.9 &	 0.27 &	 11.7 &	 21.1 &	 1017.8 &	 288.1 &	 255.4 &	$-7.57$ &	$-8.86$ &	 11.0 &	   7.2                                                         \\
Apr &	 32.5 &	  2.5 &	 30.0 &	 2.8 &	 1.4 &	 3.9 &	 0.24 &	 12.2 &	 21.0 &	 1016.7 &	 288.3 &	 255.2 &	$-6.53$ &	$-8.01$ &	 11.1 &	   7.7                                                         \\
May &	 32.5 &	  7.5 &	 25.0 &	 3.1 &	 1.8 &	 4.1 &	 0.19 &	 14.4 &	 21.1 &	 1015.6 &	 291.3 &	 257.7 &	$-5.00$ &	$-6.70$ &	  8.3 &	   5.1                                                         \\
Jun &	 35.0 &	 15.0 &	 20.0 &	 2.6 &	 2.2 &	 2.9 &	 0.07 &	 16.3 &	 21.3 &	 1015.1 &	 292.5 &	 259.6 &	$-4.38$ &	$-2.54$ &	  8.0 &	   3.4                                                         \\
Jul &	 37.5 &	 17.5 &	 20.0 &	 2.7 &	 2.1 &	 3.2 &	 0.10 &	 18.2 &	 21.3 &	 1014.8 &	 294.7 &	 260.8 &	$-4.09$ &	$-4.14$ &	  5.8 &	   2.0                                                         \\
Aug &	 40.0 &	 17.5 &	 22.5 &	 2.9 &	 2.0 &	 3.5 &	 0.13 &	 17.6 &	 21.1 &	 1014.9 &	 294.8 &	 260.3 &	$-4.43$ &	$-4.85$ &	  5.7 &	   2.6                                                         \\
Sep &	 40.0 &	 15.0 &	 25.0 &	 3.1 &	 2.1 &	 3.8 &	 0.14 &	 15.0 &	 20.9 &	 1016.5 &	 291.9 &	 258.1 &	$-5.80$ &	$-5.08$ &	  8.3 &	   4.4                                                         \\
Oct &	 37.5 &	 10.0 &	 27.5 &	 3.5 &	 2.1 &	 4.7 &	 0.20 &	 13.3 &	 21.4 &	 1018.4 &	 288.6 &	 255.6 &	$-7.45$ &	$-6.85$ &	 10.7 &	   6.7                                                         \\
Nov &	 35.0 &	  5.0 &	 30.0 &	 3.5 &	 1.9 &	 4.7 &	 0.21 &	 11.7 &	 20.7 &	 1020.1 &	 285.5 &	 253.8 &	$-9.63$ &	$-6.93$ &	 13.3 &	   8.7                                                         \\
Dec &	 30.0 &	  0.0 &	 30.0 &	 3.1 &	 1.5 &	 4.4 &	 0.24 &	 12.1 &	 21.7 &	 1020.0 &	 287.1 &	 255.6 &	$-9.71$ &	$-8.12$ &	 11.4 &	   6.9                                                         \\
Ann &	 32.5 &	  5.0 &	 27.5 &	 3.4 &	 2.0 &	 4.7 &	 0.20 &	 14.1 &	 22.1 &	 1017.0 &	 290.0 &	 257.1 &	$-6.11$ &	$-7.41$ &	  9.0 &	   5.4                                                         \\
\bottomrule
\end{tabular}
\end{center}
\end{table}

\begin{table}[!h]\footnotesize
\caption{\label{tab2}Slope values $C$ and squared correlation coefficients $R^2$ (in braces) for Reduced Major Axis regressions of
$\Delta T_s$ on $\Delta T_a$ ($\Delta T_{si} = C\Delta T_{ai}$), $\Delta T_{si} \equiv T_{si} - \overline{T_s}$,
$\Delta T_{ai} \equiv T_{ai} - \overline{T_a}$
where $X_i$ ($X = T_a,\,T_s$) is the value of $X$ in the $i$-th gridpoint
and $\overline{X}$ is the tropical mean value of $X$ (between 27.5$\degree$S and 27.5$\degree$N as in \citep{bayr13})
in a given month or annually averaged. $N$ is the number of gridpoints analyzed.}
\begin{center}
\begin{tabular}{llll}
\toprule
&Total tropics ($N = 3312$) &Ocean ($N = 2476$) & Land ($N = 836$)           \\
\cmidrule(r){1-4}
Jan				&   2.1	 (0.49)		&   1.5	 (0.57)		&   2.5	 (0.71)		 \\
Feb				&   2.0	 (0.49)		&   1.5	 (0.63)		&   2.4	 (0.64)		 \\
Mar				&   1.9	 (0.47)		&   1.5	 (0.69)		&   2.6	 (0.47)		 \\
Apr				&   1.8	 (0.48)		&   1.4	 (0.72)		&   2.8	 (0.43)		 \\
May				&   1.8	 (0.55)		&   1.3	 (0.69)		&   2.6	 (0.62)		 \\
Jun				&   1.8	 (0.60)		&   1.4	 (0.67)		&   2.4	 (0.73)		 \\
Jul				&   1.8	 (0.62)		&   1.5	 (0.68)		&   2.2	 (0.75)		 \\
Aug				&   1.8	 (0.61)		&   1.6	 (0.67)		&   2.2	 (0.69)		 \\
Sep				&   1.8	 (0.56)		&   1.6	 (0.64)		&   2.3	 (0.53)		 \\
Oct				&   1.8	 (0.46)		&   1.6	 (0.62)		&   2.8	 (0.30)		 \\
Nov				&   1.9	 (0.39)		&   1.6	 (0.59)		&   2.9	 (0.36)		 \\
Dec				&   2.1	 (0.44)		&   1.5	 (0.57)		&   2.7	 (0.61)		 \\
Annual	&   1.8	 (0.44)	 &   1.5	 (0.65)	 &   2.8	 (0.46)		 \\
\bottomrule
\end{tabular}
\end{center}
\end{table}

\bibliographystyle{ametsoc}
\bibliography{met-refs}

\end{document}